\documentclass[preprint,amsmath,amssymb,aps]{revtex4-1}
\usepackage{graphicx}
\usepackage{dcolumn}
\usepackage{booktabs}
\usepackage{bm}
\usepackage{subfig}
\usepackage{amsmath}

\usepackage{color}

\graphicspath{{FIGNew/}}

\begin{document}

\preprint{APS/123-QED}

\title{A high-order lattice Boltzmann model for the Cahn-Hilliard equation}

\author{Chunhua Zhang}
\affiliation{
State Key Laboratory of Coal Combustion, Huazhong University of Science and Technology, Wuhan 430074, China}
\author{Zhaoli Guo}
\email{zlguo@hust.edu.cn}
\affiliation{
State Key Laboratory of Coal Combustion, Huazhong University of Science and Technology, Wuhan 430074, China}
\author{Hong Liang}
\affiliation{
Department of Physics, Hangzhou Dianzi University, Hangzhou 310018, China}
\date{\today}

\begin{abstract}
In this paper, a lattice Boltzmann model with the single-relaxation-time model for the Cahn-Hilliard equation (CHE) is proposed. The discrete source term is redesigned through a third-order Chapman-Enskog analysis. By coupling the Navier-Stokes equations, the time-derivative term in the source term is expressed as the relevant spatial derivatives. Furthermore, the source term on the diffusive time scale is also proposed though recovering the macroscopic CHE to third order. The model is tested by simulating diagonal motion of a circular interface, Zalesak's disk rotation, circular interface in a shear flow, a deformation field and the problem of Rayleigh-Taylor instability. It is shown that the proposed method can track the interface with high accuracy and stability. For the complex flow, the source term on the diffusive time scale should be considered for capturing the interface correctly.
\end{abstract}
\maketitle

\section{Introduction}
Two phase flows with complex interfacial dynamics  appear in many fields of science and engineering applications. It is important to develop effective and accurate numerical methods for simulating such flows. In the designing of numerical methods, it is critical to describe the motion of fluid interface accurately. Generally, the existing numerical methods for  multiphase flows can be divided into two categories, i.e. sharp interface methods~\cite{hirt1981volume,sussman1999adaptive} and diffuse interface methods~\cite{anderson1998diffuse,jacqmin1999calculation,badalassi2003computation,ding2007diffuse,zheng2005lattice}.

In a sharp interface method, the fluid interface is treated as a sharp discontinuity with zero thickness that separates the two fluids. The hydrodynamics of each fluid is described by individual governing equations which can be solved by standard numerical techniques. In such methods the interface just serves as a moving boundary with compatible conditions through which the effects of interfacial properties on the flow are incorporated. Therefore, it is critical to capture accurately the change and motion of the interface in sharp-interface methods. On the other hand, in a diffuse interface method, the interface is replaced  by a transition region of small but finite width, across which density, viscosity, and other physical quantities of the two-phase fluid  vary smoothly.
The hydrodynamics of the whole system is described by a single set of governing equations (Navier-Stokes equations) with a body force term accounting for the interfacial force, which can be modeled based on surface tension (continuum surface force model) or fluid free-energy (phase-field model). In the latter case, the motion and topological change of the interface are usually described by the evolution of an order parameter governed by a phase-field equation, such as the Cahn-Hilliard equation (CHE)~\cite{cahn1958free,gurtin1996generalized,anderson1998diffuse} or Allen-Cahn equation (ACE)~\cite{allen1979microscopic,shen2010numerical,chiu2011conservative}.

Diffuse interface methods have some advantages in simulating interface movement and deformation on fixed grids. Particularly, numerical methods based on phase-filed models have attracted much interest in recent years, among which the lattice Boltzmann equation (LBE) method has gained much success due to its simplicity and efficiency~\cite{inamuro2004lattice,lee2005stable,lee2010lattice,zheng2006hw,liang2014phase,yang2016lattice,
shao2014free}.
Generally, two sets of LBE's are used in phase-field LBE models, one is employed to solve the phase-field equation (CHE or ACE) and the other for the hydrodynamic equations. The LBE for the Navier-Stokes equations is standard, but the LBE for the phase-field equation is nontrivial, and a number of models have been developed for both CHE and ACE. The first attempt to device a LBE describing the evolution of a phase-field varable is due to He \emph{et al.}~\cite{he1999lattice}, which reproduces an equation similar to the CHE but with some explicit differences. Later, Zheng \emph{et al.}~\cite{zheng2005lattice} proposed a modified version with a spatial term of the distribution function such that the CHE can be recovered exactly. Following the same idea, Zu \emph{et al.}~\cite{zu2013phase} further simplified the model by replacing the distribution function with the equilibrium one. However, the numerical stability of these models are very sensitive to the choice of relaxation time, more specifically, the models become unstable as the relaxation time approaches to 1. Recently, Liang \emph{et al.}~\cite{liang2014phase} developed a LBE model by introducing a time-derivative source term to recover the
CHE exactly. Some LBE models for the Allen-Cahn equation have also been developed ~\cite{geier2015conservative,ren2016improved,fakhari2010phase,wang2016comparative}. For instance,
Geier \emph{et al.}~\cite{geier2015conservative} developed a central-moment LBE model for the ACE, and Fakhari \emph{et al.}~\cite{fakhari2010phase}
employed a finite-difference LBE model for the ACE to facilitate the use of non-uniform grids. However,
it is found that both models cannot recover the ACE exactly, and a model with a time-derivative source term was proposed to overcome this problem~\cite{wang2016comparative}.

All of the existing LBE models for the CHE and ACE are based on second-order Chapman-Enskog analysis. As indicated in~\cite{zhai2017pseudopotential,huang2016third}, the high-order effects are necessary for the pseudo-potential LBE. This suggests that
the high-order effects may be also important for the LBE for the phase-field equation. In this work, we aim to propose a LBE which can match the CHE up to third-order in terms of the Chapman-Enskog analysis.

The rest of this paper is organized as follows. In Sec. II, the LBE model for the CHE is introduced with the third-order Chapman-Enskog analysis, from which the source term is determined. In Sec. III, some numerical simulations are carried out to validate the proposed model, with some comparisons with recent models. A brief summary is presented in Sec. IV finally.

\section{METHODOLOGY}
\subsection{Chan-Hilliard equation}
In  phase-field theory for a two phase system (denoted by A and B, respectively), an order parameter $\phi(\bm x,t)$ is used to identify different fluid phases, e.g., $\phi=\phi_k$ denotes the bulk phase $k$. The order parameter is closely related to free energy of the system.  For an isothermal binary fluid system, the free energy can be expressed,
\begin{equation}\label{freeEnergy}
F(\phi)=\int_{V}\left[f(\phi)+\frac{\kappa}{2}|\nabla\phi|^2\right]dV,
\end{equation}
where $f(\phi)$ is the bulk free-energy density, $\kappa$ is a parameter related to surface tension, and $V$ is the control volume.
In general, the bulk free-energy density can be modeled as a double-well potential for  pseudo-vander Waals fluids \cite{jacqmin1999calculation,jacqmin2000contact},
\begin{equation}\label{bulk}
f(\phi)=\beta(\phi-\phi_A)^2(\phi-\phi_B)^2,
\end{equation}
where $\phi_A$ and $\phi_B$ are constants corresponding to phases $A$ and $B$, respectively, and $\beta$ is another constant. The two parameters $\kappa$ and $\beta$ are related to the interfacial thickness $W$ and the surface tension $\sigma$~\cite{jacqmin1999calculation,jacqmin2000contact},
\begin{equation}\label{betakappa}
\beta=\frac{12\sigma}{W|\phi_A-\phi_B|^4},\quad \kappa=\frac{3\sigma W}{2|\phi_A-\phi_B|^2}.
\end{equation}
From the free-energy, one can define the chemical potential $\mu $ of the system,
\begin{equation}\label{mu}
  \mu\equiv \frac{\delta F}{\delta \phi}=\frac{\partial f}{\partial \phi}-\kappa\nabla^2 \phi.
\end{equation}
At equilibrium, the chemical potential is constant and the profile of the interface can then be obtained. Particularly, for a planar interface the equilibrium distribution of the order parameter can be expressed as,
\begin{equation}\label{analytical}
\phi(z)=\frac{\phi_A+\phi_B}{2}+\frac{\phi_A-\phi_B}{2}\tanh\left(\frac{2z}{W}\right),
\end{equation}
where $z$ is the coordinate normal to the interface.

In a fluid system, change of the order parameter can be described by the CHE \cite{cahn1958free,cahn1959free},
\begin{equation}\label{CH}
  \frac{\partial \phi}{\partial t}+\nabla\cdot (\phi\bm u)=\nabla\cdot(M\nabla \mu),
\end{equation}
where $M$ is the mobility, and $\bm u$ is the fluid velocity governed by the Navier-Stokes equations \cite{ding2007diffuse,jacqmin1999calculation,li2012additional,guo2014numerical,shen2013mass},
\begin{subequations}\label{NS}
\begin{align}
&\nabla\cdot \bm u=0  \\
&\rho\left(\partial_t (\bm u)+\bm u \cdot \nabla(\bm u)\right) =-\nabla p+ \nabla \cdot \Pi+\bm F_{s}, \text{or}\\
&\partial_t (\rho \bm u)+ \nabla(\rho\bm u \bm u) =-\nabla p+ \nabla \cdot \Pi+\bm F_{s}+\frac{d\rho}{d\phi}M\nabla\cdot\nabla\mu.
\end{align}
\end{subequations}
where  $\rho$ is the density, $p $ is the pressure, $\Pi$ is the viscous stress tensor, i.e, $\Pi=\nu(\nabla\bm u+\nabla \bm u^T)$, with $\nu$  being the dynamic viscosity, and $F_{s}= -\phi\nabla \mu $ is the surface tension force.
The fluid density is determined by the order parameter $\phi$,
\begin{equation}\label{rho}
  \rho=\frac{\phi-\phi_B}{\phi_A-\phi_B}(\rho_A-\rho_B)+\rho_B=a\phi+b,
\end{equation}
with $a=(\rho_A-\rho_B)/(\phi_A-\phi_B)$ and $b=(\rho_B\phi_A-\phi_B\rho_A)/(\phi_A-\phi_B)$.


\subsection{The LBE model for the Cahn-Hilliard equation}
The LBE for the CHE can be written as
\begin{equation}\label{LB}
  g_i(\bm x+\bm c_i \delta t,t+\delta t)-g_i(\bm x, t)=
  -\frac{g_i(\bm x,t)-g_i^{eq}(\bm x,t) }{\tau_g}+\delta t\left(1-\frac{1}{2\tau_g}\right)S_i(\bm x, t),
\end{equation}
where $g_i(\bm x,\bm c_i,t)$ is the distribution function associated with discrete velocity $\bm c_i$ at position $\bm x$ and time t, $\delta t$ is the time step, $\tau $ is the nondimensional relaxation time, $g_i^{eq}$ is the equilibrium distribution function, and $S_i(\bm x,t)$ is a source term to be determined later to ensure the CHE can be recovered. In this study, we consider two dimensions problems and employing the two-dimensional nine velocity (D2Q9), in which the discrete velocities are given by $\bm c_0=(0,0), \bm c_1=-\bm c_3=(1,0)c, \bm c_2=-\bm c_4=(0,1)c, \bm c_5=-\bm c_7=(1,1)c$, and $\bm c_6=-\bm c_8=(-1,1)c$, with $c=\delta_x/\delta_t$ being the lattice speed ($\delta_x$ is the lattice spacing).
The equilibrium distribution function $g_i^{eq}$ in Eq.~(\ref{LB}) is defined as~\cite{huang2009mobility,liang2014phase}
\begin{equation}\label{geq}
  g_i^{eq}=\begin{cases}
  \phi+(\omega_i-1)\eta\mu- \omega_0 \phi \frac{\bm u^2}{2c_s^2} & \text {$i=0$},\\
  \omega_i\eta\mu+\omega_i \phi \left(\frac{\bm c_i \cdot \bm u}{c_s^2}+\frac{(\bm c_i\cdot \bm u)^2}{2c_s^4}-\frac{\bm u^2}{2c_s^2}\right)  & \text {$i \neq 0$},
  \end{cases}
\end{equation}
where $\eta$ is an adjustable parameter that controls the mobility, and the weights are given by $\omega_0=4/9, \omega_{1-4}=1/9$, and $\omega_{5-8}=1/36$. $c_s$ is the sound speed, which is defined as $c/\sqrt{3}$ for D2Q9.
It is easy to verify that the equilibrium distribution functions satisfy the following conditions,
\begin{equation}\label{feqmoment}
  \sum_i g_i^{eq}=\phi, \qquad \sum_i \bm c_i g_i^{eq}=\phi\bm u,\qquad
  \sum_i \bm c_i \bm c_i g_i^{eq}=c_s^2\eta\mu+\phi\bm u\bm u .
\end{equation}
The source  term is required to meet the following constraints,
\begin{equation}\label{conservation}
\quad \sum_i S_i=0, \quad \sum_i \bm c_i S_i=\bm F,
\end{equation}
where $\bm F$ is related to the fluid velocity $\bm u$ to be determined later. It is obvious that $S_i$ defined below satisfies the above constraints:
\begin{equation}\label{liangFi}
 S_i=\frac{w_i\bm c_i}{c_s^2}\cdot \bm F.
\end{equation}

Finally, the order parameter is computed from the distribution functions as follows
 \begin{equation}\label{macroscopicg}
\phi=\sum g_i.
\end{equation}

Now we make a third-order Chapman-Enskog analysis of the above LBE model to determine the source term. The idea behind the Chapman-Enskog analysis is that different physical phenomena happen at different time scales. Usually,
a second-order analysis involving convetive and diffusive time scales is made for LBE, but here we will perform a
third-order analysis such that the CHE can be recovered more accurately. In the Chapman-Enskog analysis, the following multiscale expansions are introduced,
\begin{equation}\label{expanded}
\begin{aligned}
g_i&=g_i^{(0)}+\epsilon g_i^{(1)}+\epsilon^2 g_i^{(2)}+\epsilon^3 g_i^{(3)}+\ldots,\qquad
S_i=\epsilon S_i^{(1)}+\epsilon^2 S_i^{(2)},\qquad
\bm F=\epsilon \bm F^{(1)}+\epsilon \bm F^{(2)},\\
\partial_t &=\epsilon \partial_{t_1}+\epsilon^2\partial_{t_2}+\epsilon^3\partial_{t_3},\qquad
\nabla=\epsilon \nabla_1,
\end{aligned}
\end{equation}
where $\epsilon$ is a small expansion parameter, and $t_1$ and $t_2$ are the convective and diffusive scales,
respectively. By making a Taylor expansion of Eq.~(\ref{LB}) and substituting these expansions, we can obtain the following equations at different orders of $\epsilon$,
\begin{subequations}\label{CE}
\begin{align}
g_i^{(0)}=g_i^{eq},\\
D_{1i} g_i^{(0)} =
 - \frac{1}{\tau_g} g_i^{(1)} +\left(1-\frac{1}{2\tau_g}\right) S_i^{(1)},\\
 \partial_{t_2} g^{(0)}+D_{1i} g_i^{(1)}+\frac{1}{2}D_{1i}^2 g^{(0)}
 = - \frac{1}{\tau_g} g_i^{(2)} +\left(1-\frac{1}{2\tau_g}\right) S_i^{(2)},\\
 \partial_{t_3} g_i^{(0)}+\partial_{t_2}g_i^{(1)}+\partial_{t_2}D_{1i} g^{(0)}+
 D_{1i} g_i^{(2)}+\frac{1}{2}D_{1i}^2 g_i^{(1)}+\frac{1}{6}D_{1i}^3 g_i^{(0)}
   =-\frac{1}{\tau_g}g_i^{(3)},
\end{align}
\end{subequations}
where $D_{1i}=\partial_{t_1}+\bm c_i\cdot \nabla_1$.
In the following we will frequently use the following results from the conservation of phase field parameter and the constraints on the source terms,
\begin{equation}\label{constraints}
\sum_{i} g_i^{(n)}=0,\quad \sum_{i}F_i^{(n)}=0,\quad \sum_{i} \bm c_i S_i^{(n)}=\bm F^{(n)},
\end{equation}
for $n\geq1$. With Eqs.~(\ref{feqmoment}) and (\ref{constraints}), taking zeroth-order moment of Eq.~(\ref{CE}) gives
\begin{subequations}\label{moments0order}
\begin{align}\label{eq:dt1Phi}
\partial_{t_1}\phi+ \nabla_1 \cdot (\phi \bm u) & =0,\\
\partial_{t_2}\phi-\left(\tau_g-\frac{1}{2}\right)\sum_i D_{1i}^2 g_i^{(0)}
+\left(\tau_g-\frac{1}{2}\right)\sum_i D_{1i}S_i^{(1)} &=0,\\
\partial_{t_3}\phi+\frac{1}{2}\sum_i D_{1i}^2 g_i^{(1)}+ \frac{1}{6}\sum_i  D_{1i}^3 g_i^{(0)}
+\sum_i  D_{1i} g_i^{(2)} &=0.
\end{align}
\end{subequations}
By substituting Eq.~(\ref{CE}b) into Eq.~(\ref{moments0order}b), we have
\begin{equation}\label{phit2}
\partial_{t_2}\phi -\left(\tau_g-\frac{1}{2}\right)\sum_i {\left[D_{1i}^2 g_i^{(0)}- D_{1i} S_i^{(1)} \right]}=0.
\end{equation}
Combining Eqs.~(\ref{moments0order}a) and (\ref{phit2}) leads to
\begin{equation}\label{ch-equation-t2}
\partial_t \phi + \nabla_1 \cdot (\phi\bm{u})= \left(\tau_g-\frac{1}{2}\right) \epsilon \nabla_1 \cdot \left[ \epsilon \partial_{t_1} (\phi \bm u)+ \epsilon \nabla_1\cdot (c_s^2 \eta \mu+\phi\bm u\bm u)- \epsilon F^{(1)} \right]+O(\epsilon^3).
\end{equation}
In order to recover the CHE up to the order of $\epsilon^2$, we must choose
\begin{equation}\label{cFi1}
 \bm F^{(1)} =\partial_{t_1} (\phi\bm u)+\nabla_1 \cdot (\phi\bm u\bm u).
\end{equation}

With the help of the first-order incompressible Navier-Stokes equation in $\epsilon$~\cite{chai2013lattice,chai2014nonequilibrium},  we can obtain the equation on the $t_1$ time scale,
\begin{equation}\label{NS-t00}
\rho(\partial_{t_1}\bm u+\bm u \cdot \nabla_1  \bm u)=\bm F_s-\nabla_1 p.
\end{equation}
which gives
\begin{equation}\label{t0-u}
\partial_{t_1} \bm u=\frac{\bm F_s-\nabla_1 p}{\rho}-\bm u\cdot \nabla_1 \bm u.
\end{equation}
Then we can obtain
\begin{equation}\label{t0-phiu}
\begin{split}
 \bm F^{(1)}
&=\bm u\partial_{t_1}\phi+\phi\partial_{t_1}\bm u+\nabla_1 \cdot(\phi\bm u\bm u)\\
&=\frac{\phi}{\rho} \left(\bm F_s-\nabla_1 p\right),
\end{split}
\end{equation}

In order to identify the source term on the diffusive $\partial_{t_2}$ time scale, we combine Eq.~(\ref{ch-equation-t2}) and Eq.~(\ref{moments0order}c), to get
\begin{equation}\label{phit3}
\begin{split}
\partial_t \phi + \nabla_1 \cdot (\phi\bm{u})
=& \epsilon^2 \nabla_1 \cdot (M\nabla_1 \mu)   \\
&\underbrace{-\frac{1}{2}\sum_i D_{1i}^2 g_i^{(1)}-\frac{1}{6} \sum_i D_{1i}^3 g_i^{(0)}-D_{1i} g_i^{(2)}}_{O(\epsilon^3)}+O(\epsilon^4),
\end{split}
\end{equation}
where $M=c_s^2\eta(\tau_g-0.5)$ is the mobility. The second line of the right-hand side of the Eq.~(\ref{phit3}) is the leading error terms for the third order of the CHE. Multiplying $D_{1i}^2$ on both sides of Eq.~(\ref{CE}b) leads to
\begin{equation}\label{phit3D1}
  D_{1i}^2 g_i^{(1)}=-\tau_g\left[ D_{1i}^3 g_i^{(0)} -\left(1-\frac{1}{2\tau_g}\right)D_{1i}^2 S_i^{(1)}\right],
\end{equation}
and multiplying $D_{1i}$ on both sides of Eq.~(\ref{CE}c) leads to
\begin{equation}\label{phit3D2}
\begin{split}
D_{1i} g_i^{(2)}
=&\left(\tau_g^2-\frac{\tau_g}{2}\right) D_{1i}^3 g_i^{(0)}\\
&-\left(\tau_g^2-\frac{\tau_g}{2}\right) D_{1i}^2 S_i^{(1)}
+\left(\tau_g-\frac{1}{2} \right)\nabla_1 \cdot \bm c_i S_i^{(2)},
\end{split}
\end{equation}
where we have used Eqs.~(\ref{moments0order}a) and (\ref{constraints}) in the derivation.
Inserting Eqs.~(\ref{phit3D1}) and (\ref{phit3D2}) into Eq.~(\ref{phit3}) leads to
 \begin{equation}\label{phit3inserting}
 \begin{split}
 \partial_t \phi + \nabla_1 \cdot (\phi\bm{u})
=&\nabla_1 \cdot (M\nabla_1 \mu)   \\
&+\underbrace{\left(\tau_g-\frac{1}{6}-\tau_g^2 \right)\sum_i D_{1i}^3 g_i^{(0)}}_{{R_1}}\\
&+\underbrace{\left(\tau_g-\frac{1}{2}\right)^2 \sum_i{(\partial_{t_1}+\bm c_i\nabla_1)^2 S_i^{(1)} }
 -\left(\tau_g-\frac{1}{2}\right)\sum_i{(\partial_{t_1}+\bm c_i\nabla_1) S_i^{(2)}}}_{{R_2}}+O(\epsilon^4).
\end{split}
\end{equation}
In order to reduce the error caused by the three order term, one way is to let $R_1$ be zero by setting $\tau_g-\frac{1}{6}-\tau_g^2=0$. Then, we can get a special relaxation time $\tau_g=0.5+\sqrt{3}/6$, which is consistent with the previous results~\cite{servan2008lattice,van2006galilean}. However, $R_2$ cannot be guaranteed to be zero. In order to completely eliminate the third-order items, $R_1+R_2$ must be zero,
 which leads to the discrete source term on the diffusive time scale,
\begin{equation}\label{F2}
\begin{split}
 \nabla\cdot \sum_i \bm c_i S_i^{(2)}
  =&\left(\tau_g-\frac{1}{6}-\tau_g^2 \right) \left(\tau_g-\frac{1}{2}\right)^{-1}\sum_i D_{1i}^3 g_i^{(0)} \\
 &+\left(\tau_g-\frac{1}{2}\right)\left(2\partial_{t_1} \nabla_1\cdot \sum_i \bm c_i S_i^{(1)}+ \nabla_1^2\sum_i \bm c_i\bm c_i S_i^{(1)}\right).
\end{split}
\end{equation}
or
\begin{equation}\label{F2-F}
\begin{split}
 \nabla\cdot \bm  F^{(2)}
  =&\left(\tau_g-\frac{1}{6}-\tau_g^2 \right) \left(\tau_g-\frac{1}{2}\right)^{-1}\sum_i D_{1i}^3 g_i^{(0)} \\
 &+\left(\tau_g-\frac{1}{2}\right)\left(2\partial_{t_1} \nabla_1\cdot \bm F^{(1)}\right).
\end{split}
\end{equation}
where we have used the fact that $\sum_i \bm c_i \bm c_i S_i^{(1)}=0$ with Eq.~(\ref{F2-F}). From Eq.~(\ref{geq}), the term $\sum_i D_{1i}^3 g_i^{(0)}$ can be expressed as
\begin{equation}\label{Di3geq}
\begin{split}
\sum_i D_{1i}^3 g_i^{(0)}&=\partial_{t_1}\partial_{t_1}\partial_{t_1}\phi +3 \partial_{t_1}\partial_{t_1}\nabla_1 \cdot(\phi\bm u)+3\partial_{t_1}\nabla_1\nabla_1:(\phi
\bm u\bm u+c_s^2\eta\mu \bm I)+3c_s^2\nabla_1^2(\nabla_1\cdot(\phi\bm u))\\
&=\partial_{t_1}\partial_{t_1}\partial_{t_1}\phi +
3\partial_{t_1}\nabla_1\cdot \bm F^{(1)}
+3c_s^2\eta \partial_{t_1}\nabla_1^2 \mu
+3c_s^2\nabla_1^2(\nabla_1\cdot(\phi\bm u)).
\end{split}
\end{equation}
Based on Eq.~(\ref{eq:dt1Phi}), $\partial_{t_1}\partial_{t_1}\partial_{t_1}\phi$ is of the third order of the Mach number.
Therefore, the above equation can be reduced to
\begin{equation}\label{Di3geq-1}
\begin{split}
\sum_i D_{1i}^3 g_i^{(0)}
&=3\partial_{t_1}\nabla_1 \bm F^{(1)}+3c_s^2\eta \partial_{t_1}\nabla_1^2 \mu +3c_s^2 \nabla_1^2(\nabla_1 \cdot(\phi \bm u)).
\end{split}
\end{equation}
Substituting Eq.(\ref{Di3geq-1}) into Eq.(\ref{F2-F}), we can write the $\bm F^{(2)}$ as
\begin{equation}\label{F2-1}
\begin{split}
 \bm F^{(2)}
&=  \left(2\tau_g-1+3K\right)\partial_{t_1}\bm F^{(1)}+3c_s^2\eta K \partial_{t_1}\nabla_1 \mu+ 3c_s^2K\nabla_1 (\nabla_1\cdot(\phi \bm u)).
\end{split}
\end{equation}
where $K=\left(\tau_g-\frac{1}{6}-\tau_g^2 \right)/\left(\tau_g-\frac{1}{2}\right)$.

Finally, combining $\bm F^{(1)}$ and $\bm F^{(2)}$, we can get the expression of $\bm F$ with second-order effect
\begin{equation}\label{reFi}
\bm F=\epsilon \bm F^{(1)}+\epsilon^2 \bm F^{(2)} = \underbrace{\frac{\phi}{\rho}(\bm F_s -\nabla_1 p)}_{\text{first-order}}+
\underbrace{ \left(2\tau_g-1+3K\right)\partial_{t_1}\bm F^{(1)}+3c_s^2\eta K \partial_{t_1}\nabla_1 \mu+3c_s^2K\nabla_1(\nabla_1 \cdot(\phi\bm u))}_{\text{second-order}}.
\end{equation}
We refer the LBE model with the source term containing $\bm F^{(2)}$ given by Eq.~(\ref{reFi}) as second-order LBE (Model-II)
while that with $\bm F$ containing only $\bm F^{(1)}$ as first-order LBE (Model-I).
Certain remarks on the source term $\bm F^{(2)}$ are given below:

Remark I: If  $\tau_g=0.5+\sqrt{3}/6$, then $K=0$. One can get a simple expression of $\bm F^{(2)}$,
\begin{equation}\label{mark1}
\bm F^{(2)} = \left(2\tau_g-1\right)\partial_{t_1}\bm F^{(1)}.
\end{equation}
The time derivative on the convection scale can be computed by using the time
 derivative that contains all the time scales without decreasing the numerical accuracy.

Remark II: If  $\tau_g=1$, then $2\tau_g-1+3K=0$. One can obtain another simple expression of $\bm F^{(2)}$,
\begin{equation}\label{mark2-1}
 \bm F^{(2)} = 3c_s^2\eta K \partial_{t_1}\nabla_1 \mu+3c_s^2K\nabla_1(\nabla_1\cdot(\phi\bm u)),
\end{equation}
or
\begin{equation}\label{mark2-2}
 \bm F^{(2)} = 3c_s^2K \partial_{t_1}(\eta \nabla_1 \mu-\nabla_1\phi),
\end{equation}
where Eq.~(\ref{eq:dt1Phi}) has been used. Analogously, the time the time derivative on the convection scale can be replaced by the whole time derivative.

Remark III:
Note that $\mu\sim \sigma\sim U_{*}$, then $\partial_{t_1} \mu\sim |U_{*}|^2 $. If the $\eta$ is less than a certain value or the Pe number is larger a certain value. The term $3c_s^2\eta K \partial_{t_1}\nabla_1 \mu$ can be of the third order of the Mach number. In addition, in the equilibrium state, $\nabla \mu=0$, then $3c_s^2\eta K\partial_{t_1}\nabla_1\mu$ can also be neglected. Thus, for $\tau_g=1$, the above expression of $\bm F^{(2)}$ can be further simplified to
\begin{equation}\label{mark3}
 \bm F^{(2)} = 3c_s^2K\nabla_1(\nabla_1\cdot(\phi\bm u)).
\end{equation}

To avoid the time derivative and save memory usage, we use Eq.~(\ref{mark3}) as a presentation.
 It is worth nothing that the term $\nabla_1(\nabla_1\cdot(\phi\bm u))$ is related to $\nabla_1\bm u$ and $\nabla_1\nabla_1\phi$. Thus, the force on the diffusive time scale should play an important role in complex deformation.

\section{Numerical Results and discussion}
In this section, several tests will be performed to validate the accuracy and robustness of the proposed two LBE models, including a diagonal motion of a circular interface, Zalesak's disk rotation, circular interface in a shear flow, a deformation field and the problem of Rayleigh-Taylor instability.
In each test case, the results will be compared with the previous LBE model in \cite{liang2014phase}. In all simulations, we take $\phi_A=-\phi_B=1$, $\rho_A/\rho_B=2$ and $\tau_g=1$ unless otherwise stated. The spatial gradients and Laplace operators are discretized with the isotropy central schemes~\cite{lee2005stable,guo2011force}.
The relative error of the order parameter is calculated by
 \begin{equation}\label{err}
||E_{\phi}||_2=\sqrt{\frac{\sum_{x,y}(\phi(\bm x,T)-\phi(\bm x,0))^2}{\sum_{x,y}\phi(\bm x,0)^2}},
\end{equation}
where  $\phi(\bm x,0)$ is the initial solution and $\phi(\bm x,T)$ is the numerical result at $T=L_{*}/U_{*}$ with $L_{*}$ and $U_{*}$ being the characteristic length   and velocity respectively.

\subsection{Diagonal translation of circular interface}
\begin{figure}[!htb]
\centering
\includegraphics[width=0.75\textwidth,trim=0 10 0 20,clip]{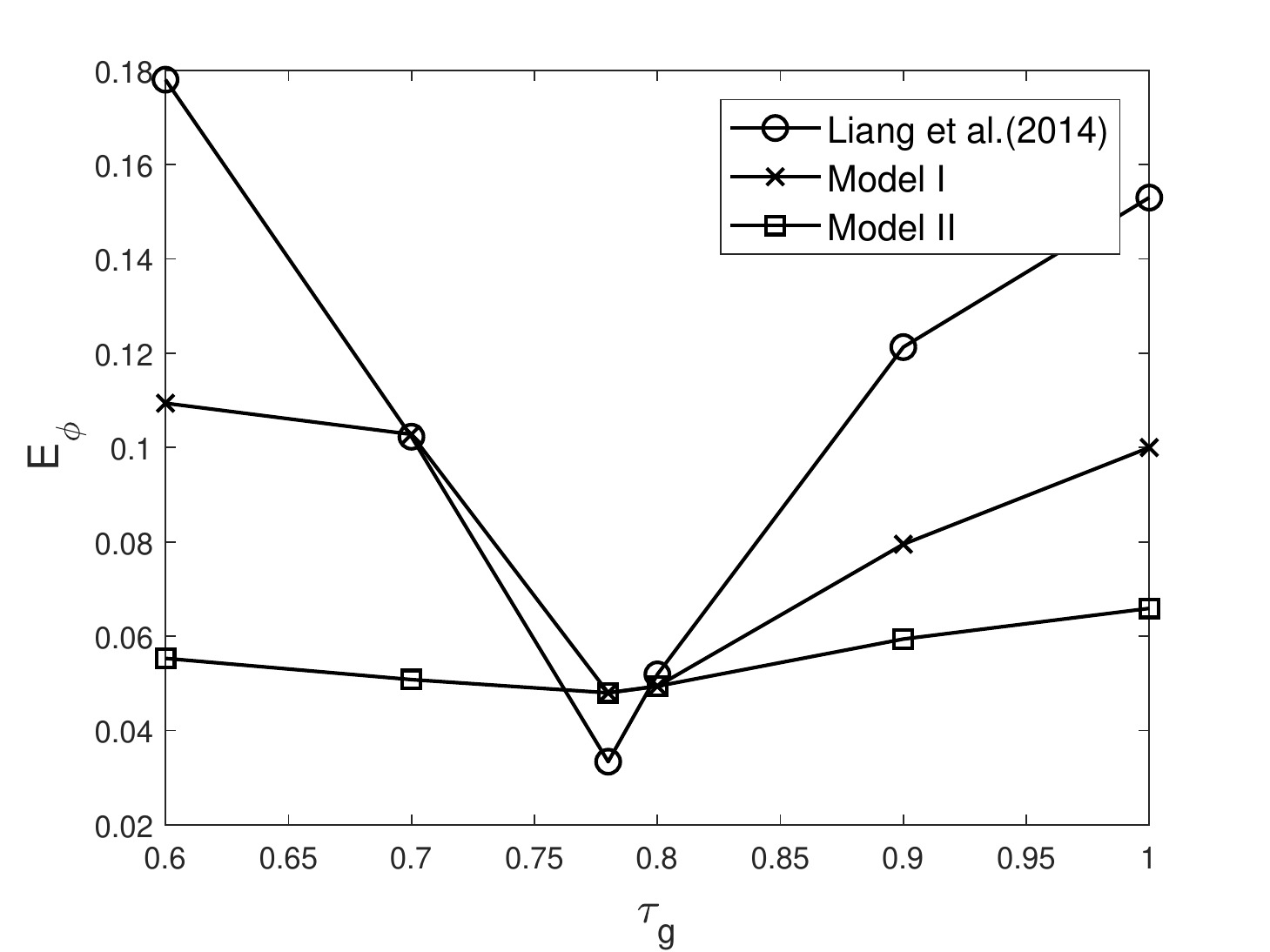}
\caption{Relative errors of the present models with different relaxation time for the diagonal translation of a circular interface}\label{fig1-0}
\end{figure}

We firstly consider a circular droplet motion under a constant velocity $\bm u=(U_0,U_0)$. Initially,
a circular droplet with radius $R=L_0/5$ is placed at the center of a periodic domain with $L_0 \times L_0$ lattice size. After a periodic $T$, the final shape should coincide with the initial shape. In simulations,
the parameters are set as follows: $L_0=200, U_0=0.02, \mbox{Pe}=200,W=2, \sigma=0.01$, and $\rho_A/\rho_B=2$.
First, we will test the relative errors of the present models and the model
of Liang \emph{et al.}~\cite{liang2014phase} with different relaxation times, as shown in Fig.~\ref{fig1-0}.
It can be found that the relative error of all models with $\tau_g=0.5+\sqrt{3}/6$ is the smallest among the parameters considered.
 Except for $\tau_g=0.5+\sqrt{3}/6$, the relative errors of model II are significantly reduced  because of considering the source term on the diffusive scale.

\begin{figure}[!htb]
\centering
\begin{tabular}{ccc}
\includegraphics[width=0.32\textwidth,trim=0 20 0 20,clip]{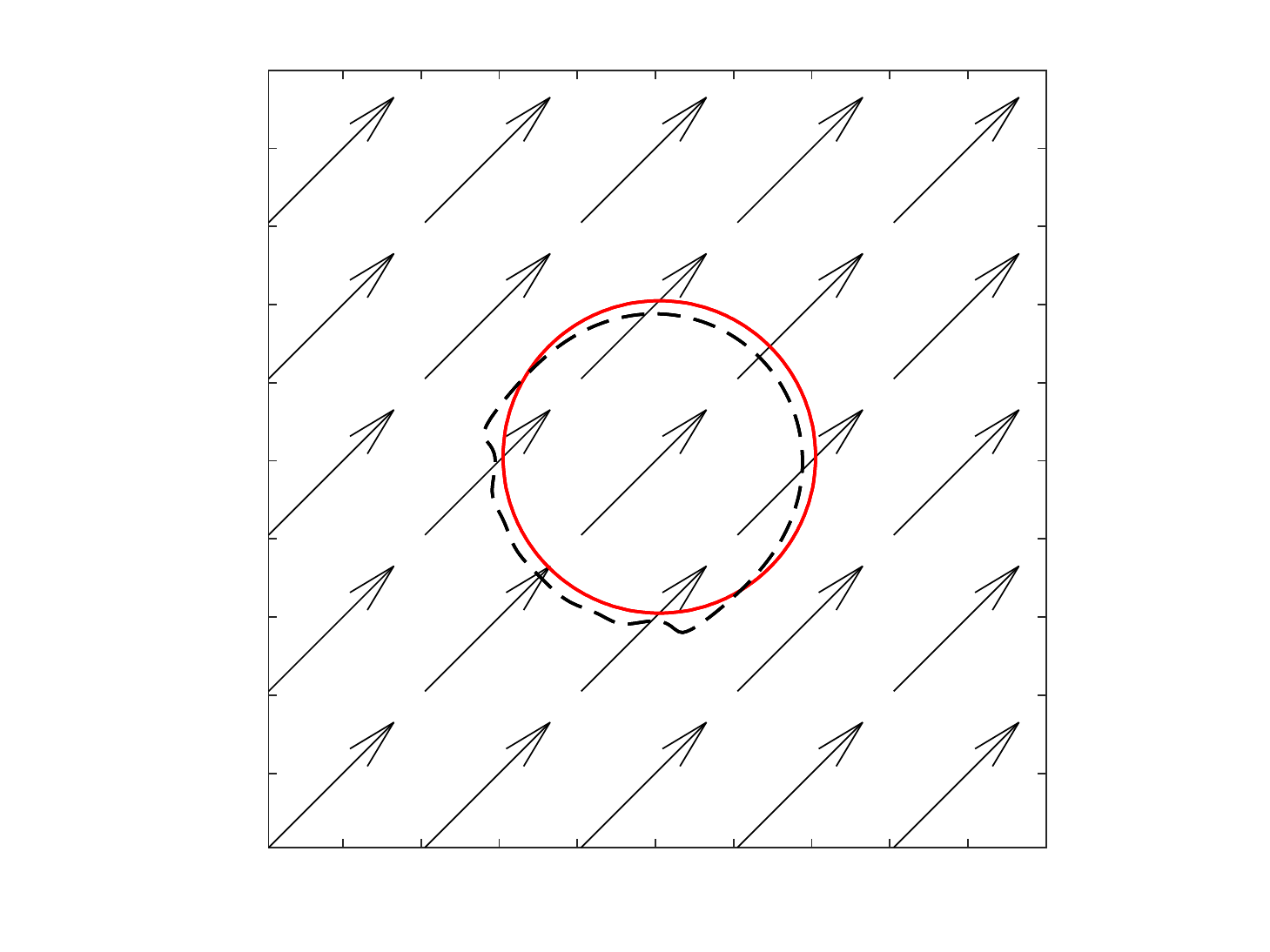}&
\includegraphics[width=0.32\textwidth,trim=0 20 0 20,clip]{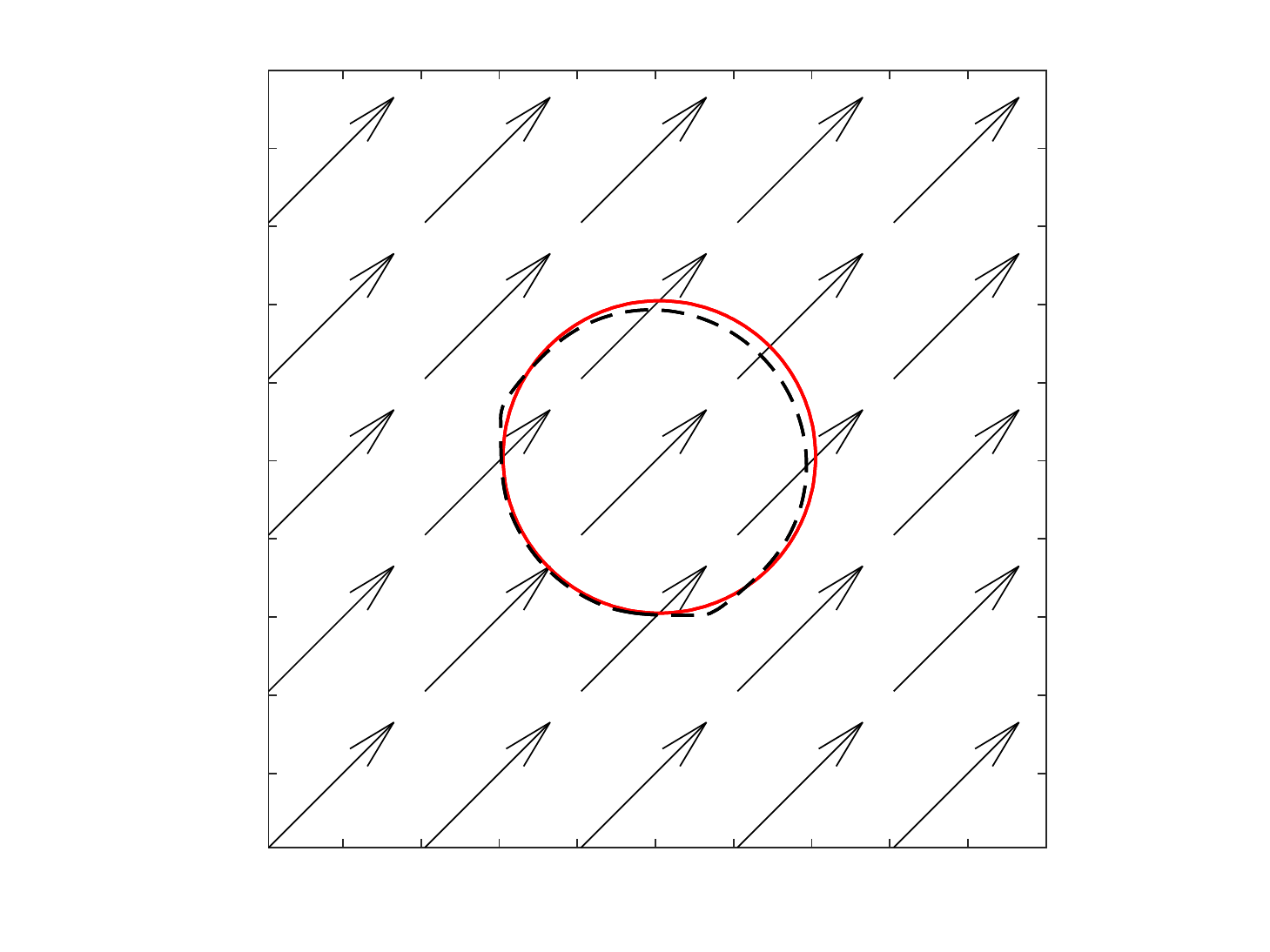}&
\includegraphics[width=0.32\textwidth,trim=0 20 0 20,clip]{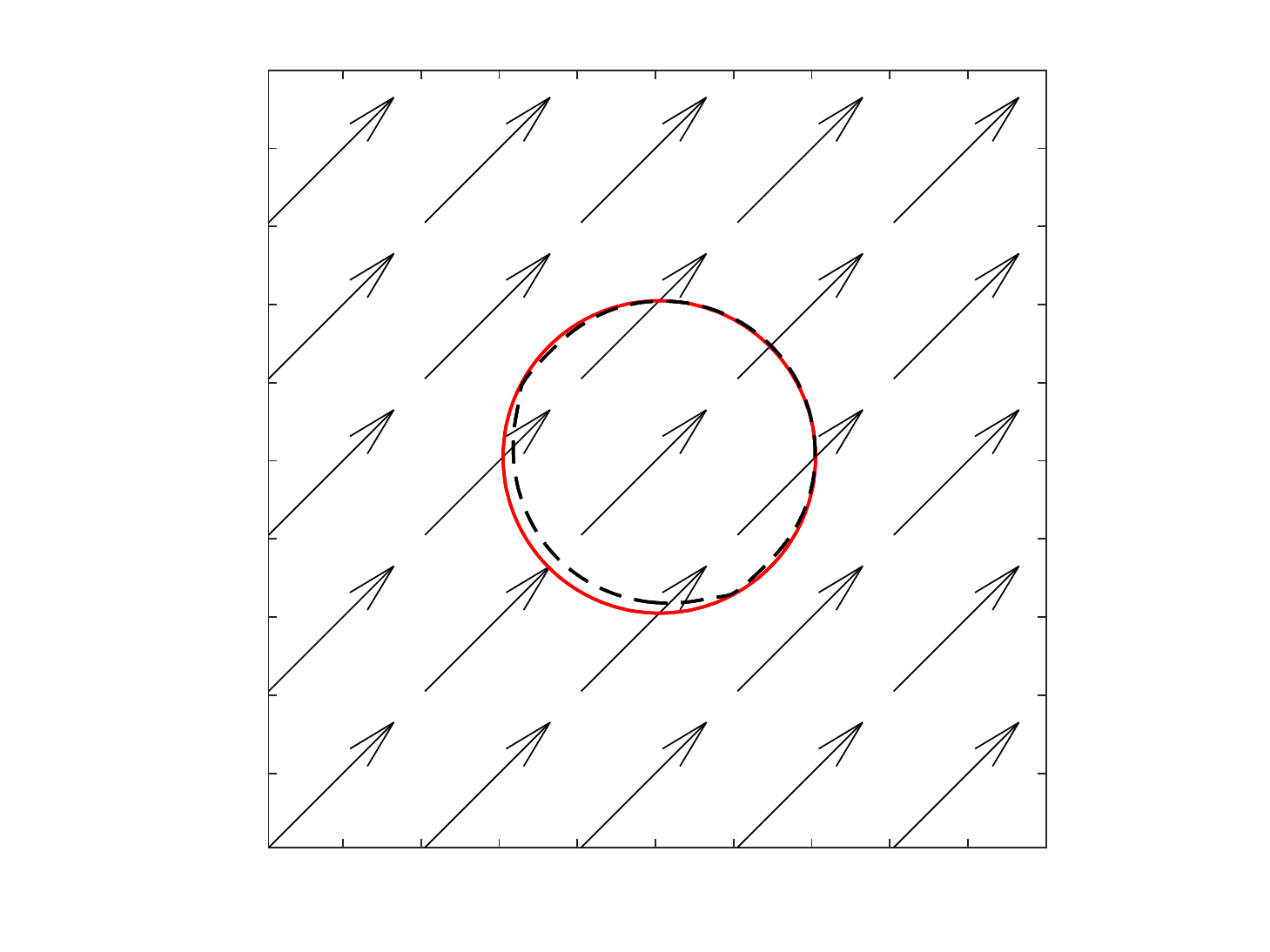}\\
(a)$\tau_g=1$&(b)$\tau_g=1,\rho_A/\rho_B=2$&(c)$\tau_g=1,\rho_A/\rho_B=2$\\
\includegraphics[width=0.32\textwidth,trim=0 20 0 20,clip]{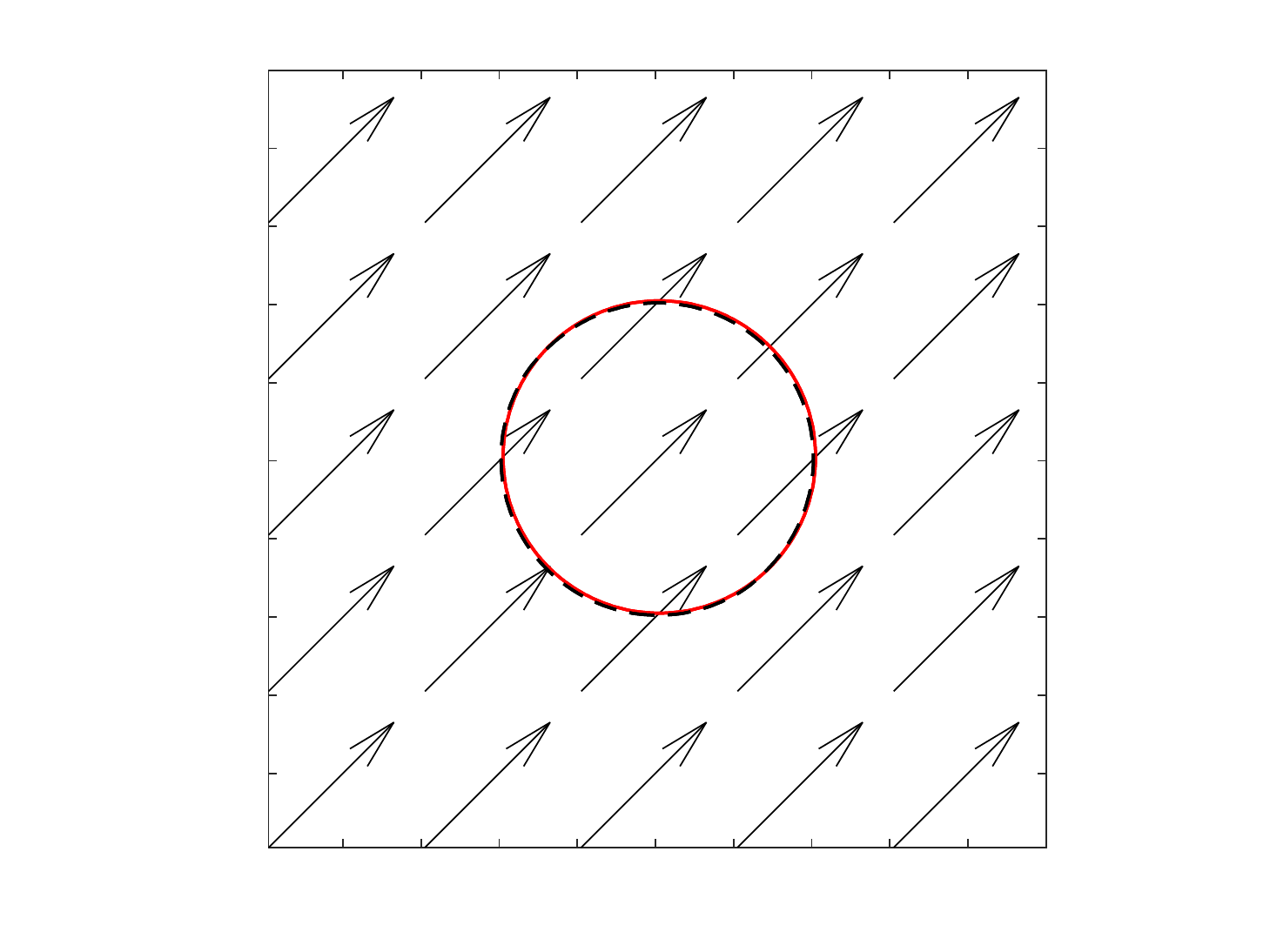}&
\includegraphics[width=0.32\textwidth,trim=0 20 0 20,clip]{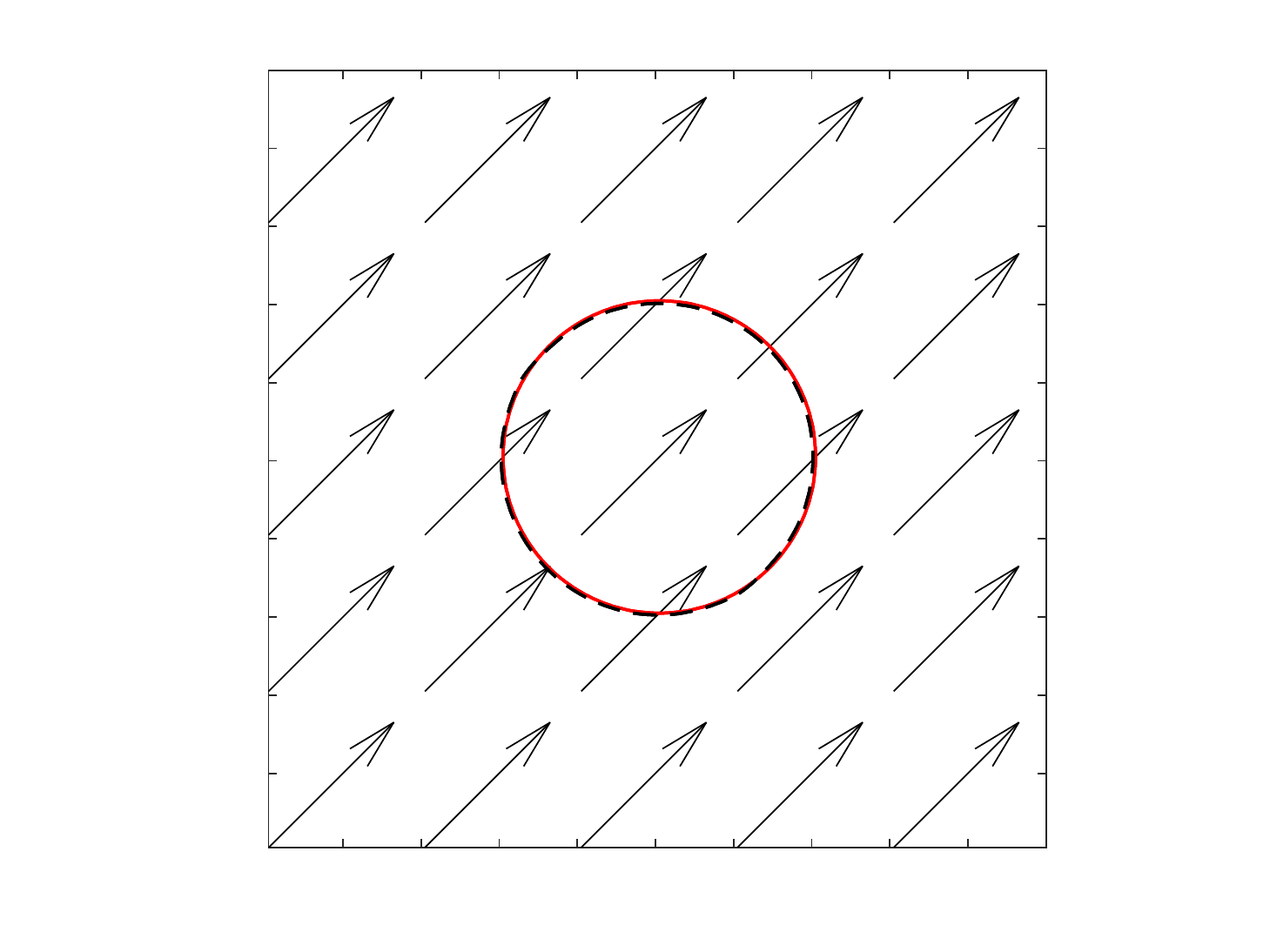}&
\includegraphics[width=0.32\textwidth,trim=0 20 0 20,clip]{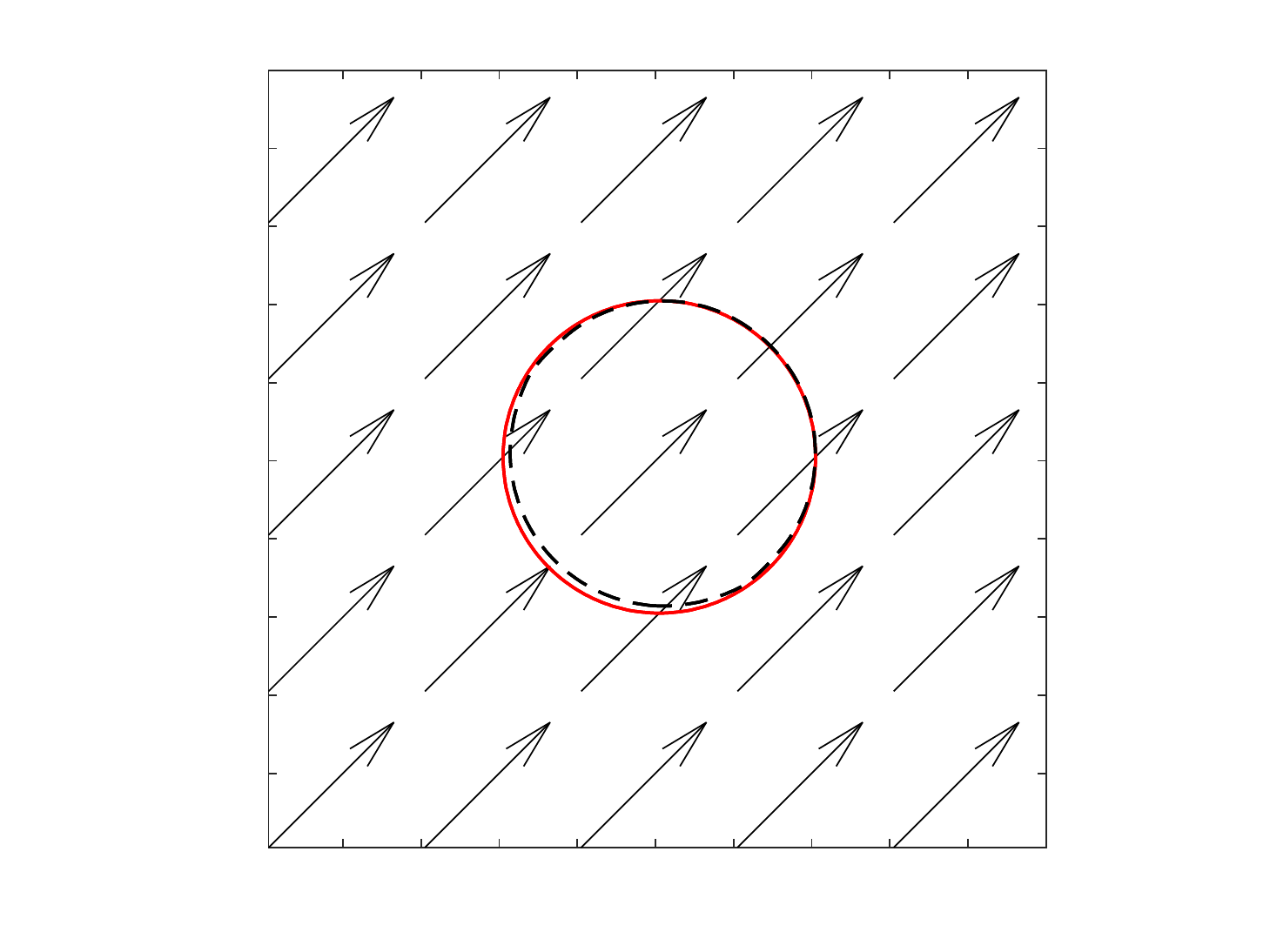}\\
(d)$\tau_g=0.5+\sqrt{3}/6$&(e)$\tau_g=0.5+\sqrt{3}/6,\rho_A/\rho_B=1$&(f)$\tau_g=0.5+\sqrt{3}/6,\rho_A/\rho_B=2$.
\end{tabular}
\caption{The phase variable contours ($\phi=0$) of diagonal motion of a circular interface at t=0 (solid line) and t=4T (dashed line) for (a)(d) the model in \cite{liang2014phase}, (b)(e)(f) model I,and (c) model II}
\label{fig1-1}
\end{figure}

Next, we present a comprehensive comparison among the present models and Liang's model~\cite{liang2014phase}. Figure~\ref{fig1-1} shows the initial profile of the interface and the final shape after 4T  at $\mbox{Pe}=200$. It is clear that the results of model II at the time $4T$ are in agreement with the initial shape, while both the model I and the model in \cite{liang2014phase} produce a slight deviation when $\tau_g$ is fixed at 1.
When  $\tau_g=0.5+\sqrt{3}/6$, the final interface profile obtained by the model in \cite{liang2014phase} is very agreement with the initial profile. On the contrary, the interface profile obtained by the model I or model II is in agreement well with  the initial profile only for a matched densities. This is due to the fact the source term in Eq.~(\ref{t0-phiu}) is closely coupled to the flow field. However, the values of the order parameter computed by both model I and model II  are closer to the initial value $\phi_A$ or $\phi_B$. To be specific, for $\tau_g=0.5+\sqrt{3}/6$ and $\rho_A=\rho_B$, the maximum and minimum values of the order parameter computed by the model in \cite{liang2014phase} are $1.2169$ and $-1.2076$, respectively, while
the maximum and minimum values of the order parameter computed by the model II are $1.0487$ and $-1.0361$, respectively.

\begin{figure}[htb]
\centering
\includegraphics[width=0.75\textwidth,trim=0 0 0 20,clip]{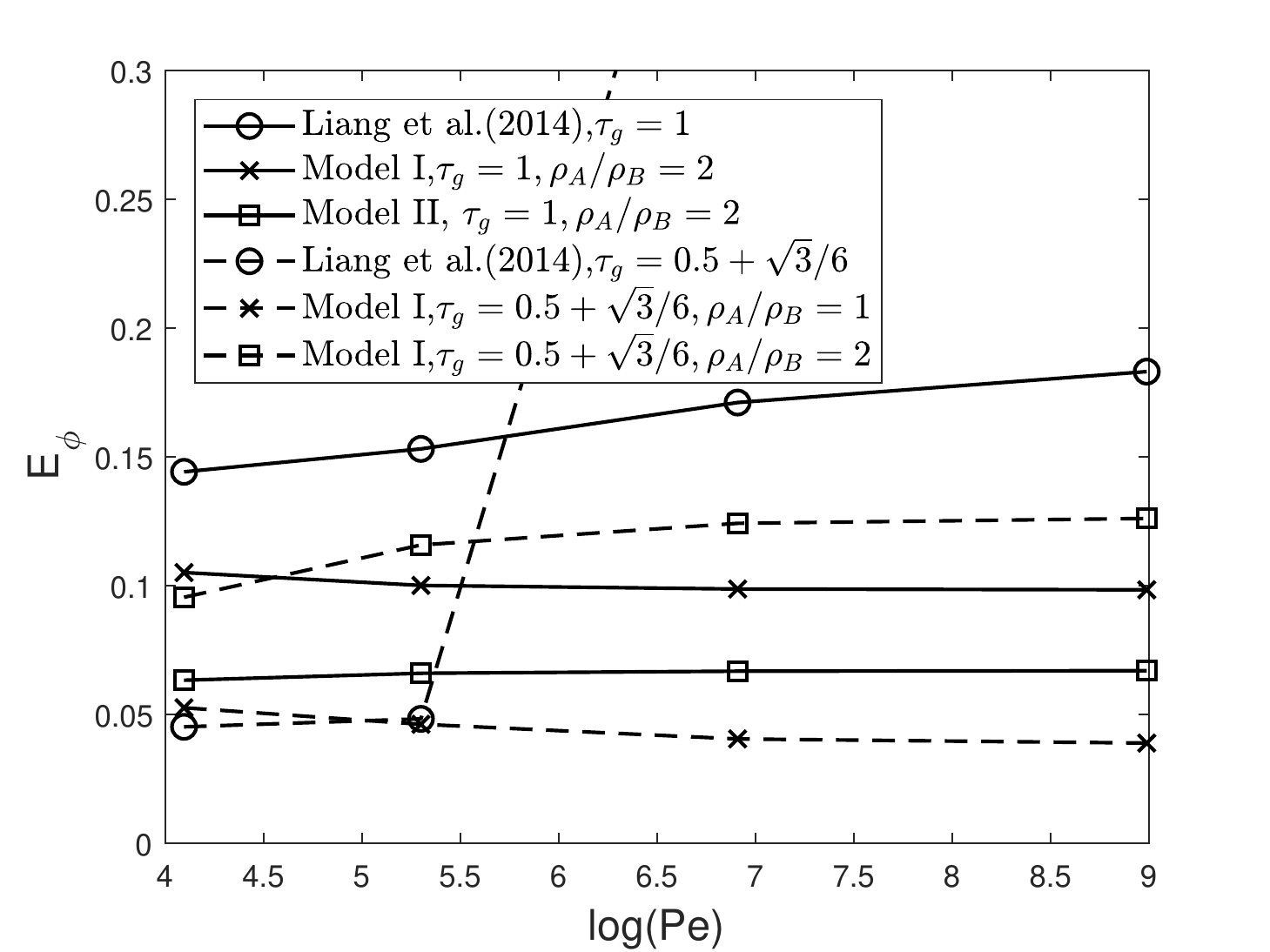}
\caption{Relative errors of different models with \mbox{Pe} numbers for the diagonal translation of a circular interface.}\label{fig1-2}
\end{figure}

 Finally, we examine the effect of Peclet number (Pe) on the numerical results.
 The dimensionless Peclet number is defined as $Pe=U_{*}W/M\beta(\phi_A-\phi_B)^2$.
 Figure~\ref{fig1-2} shows the relative errors for the above three models with different $\text{Pe}$ numbers. As seen from this figure, the relative errors of model in \cite{liang2014phase} are very susceptible to the Pe numbers. However, the results of both the model I and model II are more accurate and stable.

\subsection{Zalesak's disk rotation}
\begin{figure}[!htbp]
\centering
\subfloat[]{\includegraphics[width=0.32\textwidth,trim=0 20 0 20,clip]{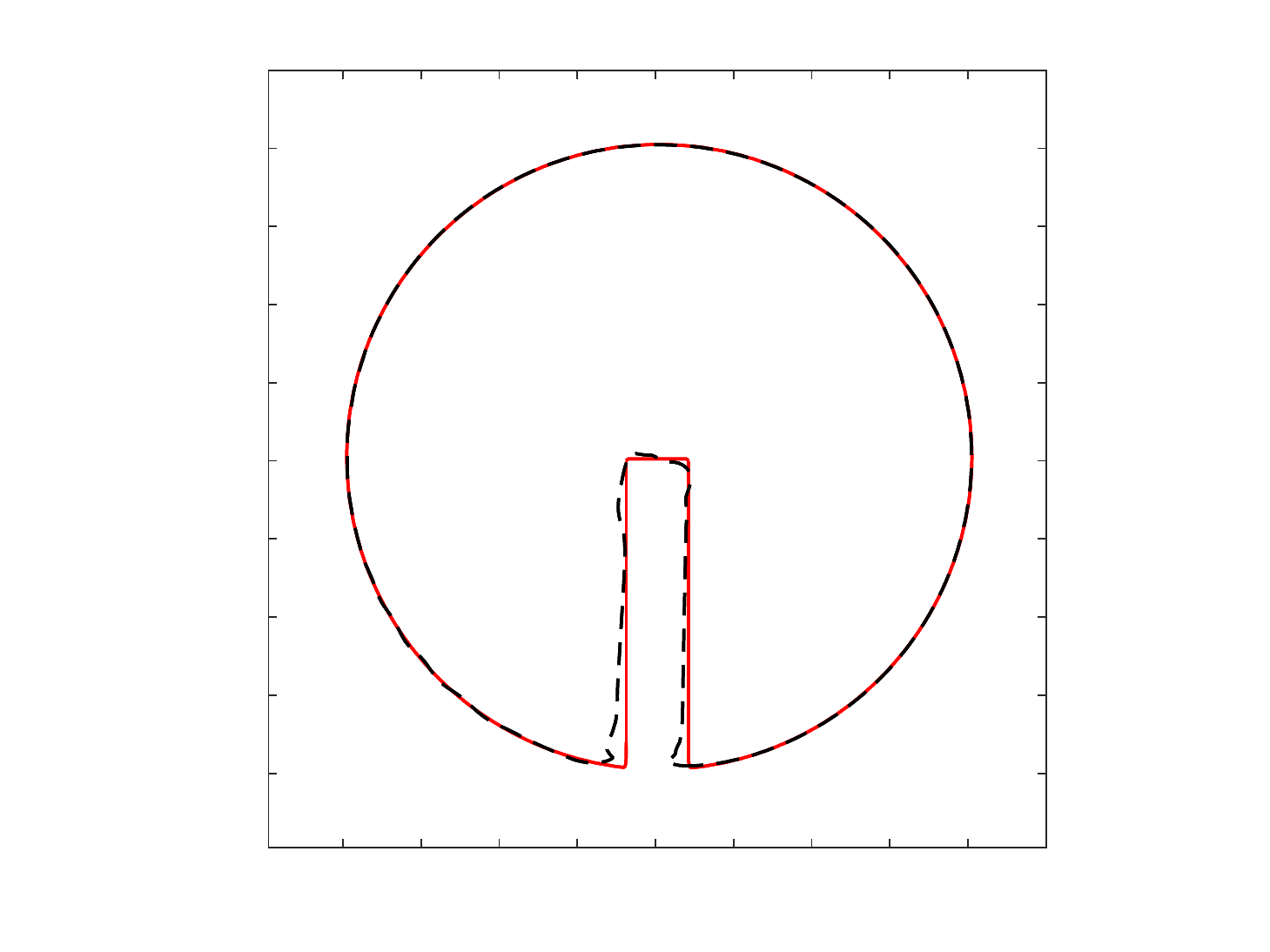}}~
\subfloat[]{\includegraphics[width=0.32\textwidth,trim=0 20 0 20,clip]{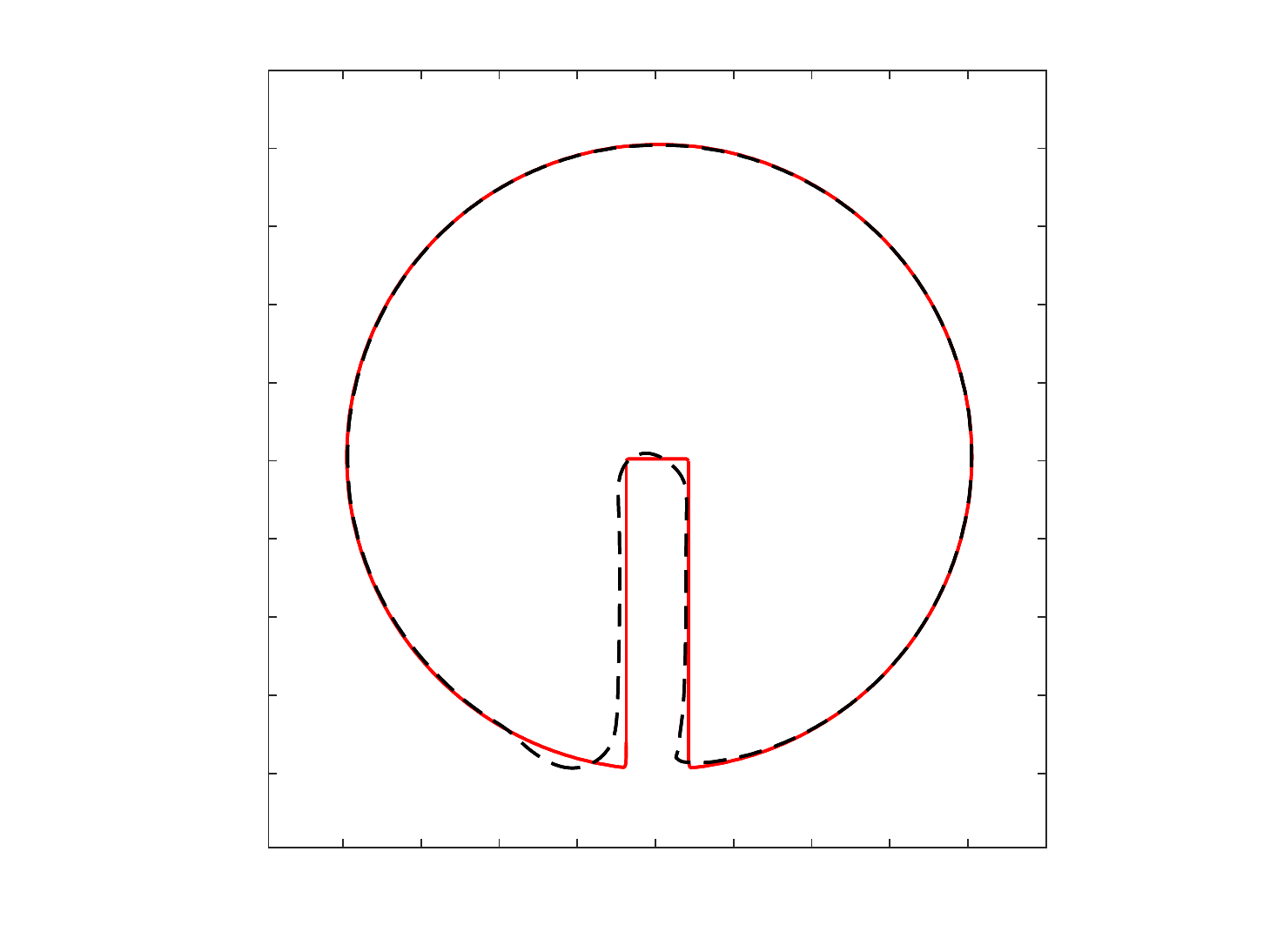}}~
\subfloat[]{\includegraphics[width=0.32\textwidth,trim=0 20 0 20,clip]{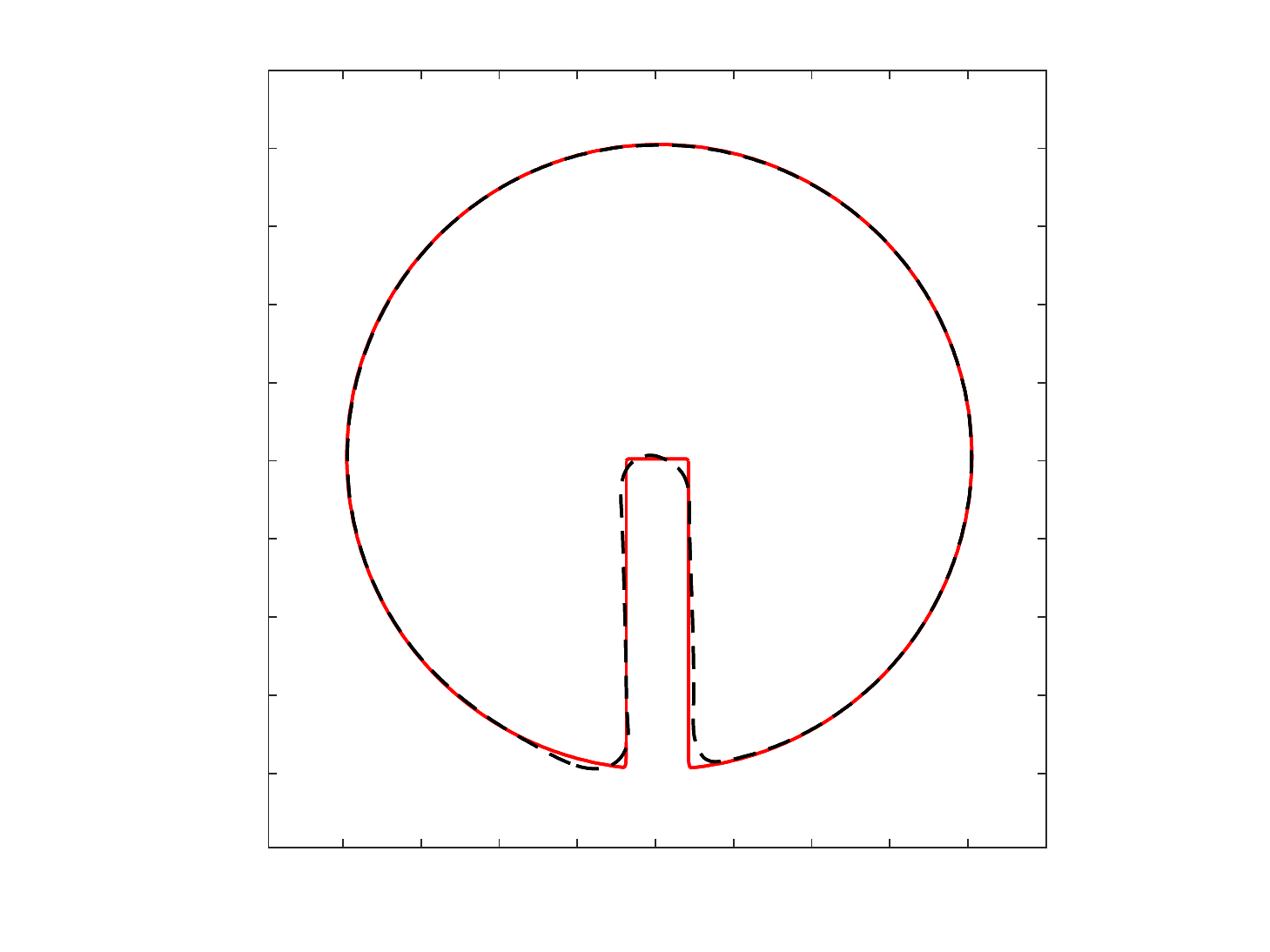}}
\caption{The phase-field contours ($\phi=0$) of Zalesak's disk at $U_0=0.02$ and $Pe=200$ for (a) the model in \cite{liang2014phase}, (b) model I, and (c) model II. The contours at $t=0$ are displayed by the solid line and the contours at $t=2T$ are displayed by the dashed line.}
\label{zalesak}
\end{figure}

\begin{figure}[!htb]
  \centering
  \includegraphics[width=0.75\textwidth,trim=0 0 0 20,clip]{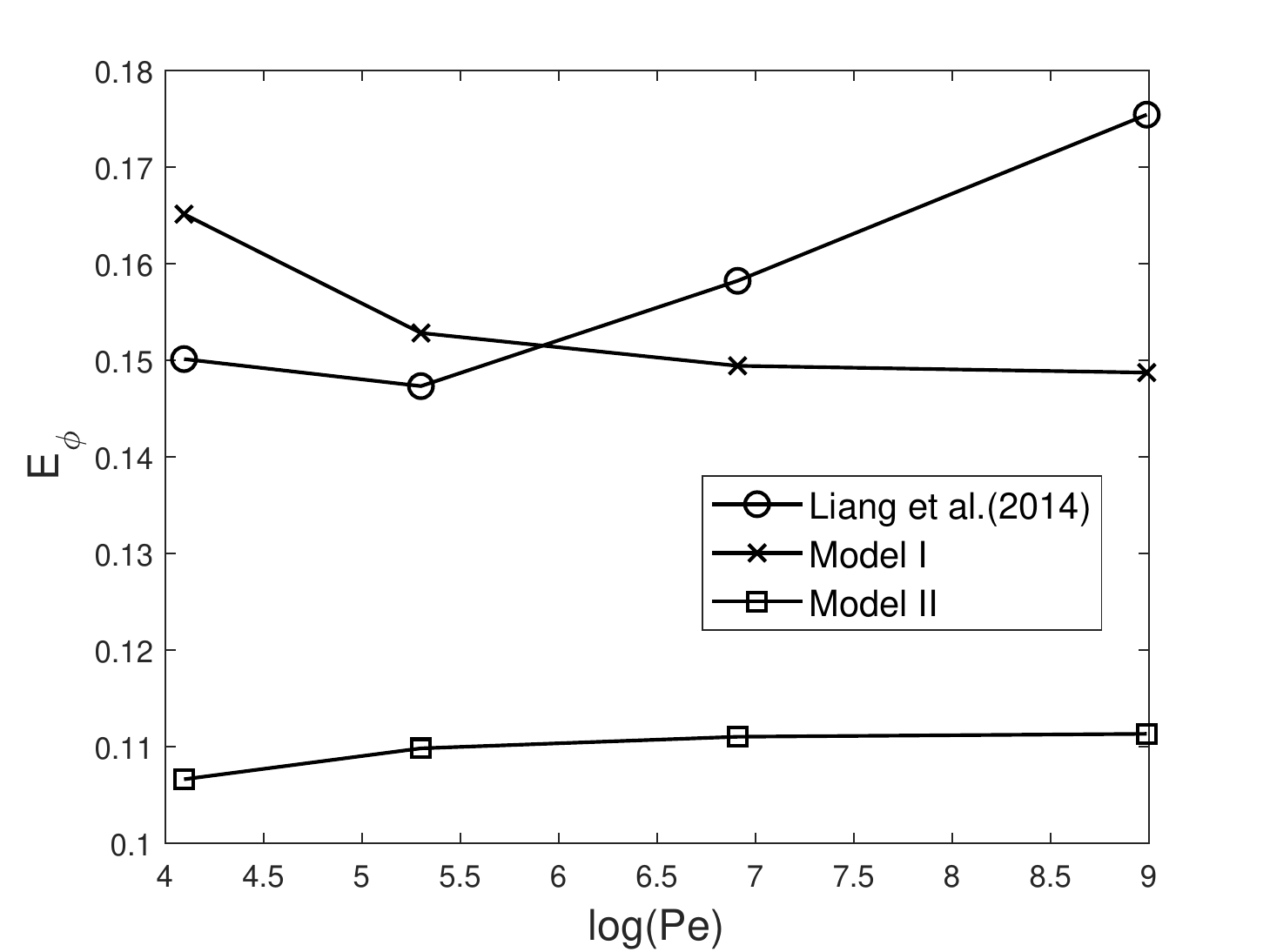}
  \caption{Relative errors of different models with $\mbox{Pe}$ numbers for Zalesak's disk rotation. }\label{fig4}
\end{figure}

The problem of Zalesak's disk is also widely used to test the capacity of the numerical methods in tracking the interface. A schematic of the problem is shown in Fig.~\ref{zalesak}. For this problem,
a circle disk with a slot is placed at the center of a periodic domain with $L_0\times L_0$ lattice size. The disk radius and the slot width are set as $80$ and $16$ lattice units, respectively.
 When the following velocity is imposed
\begin{equation}
u=-U_0\pi\left(\frac{y}{L_0}-0.5\right),\quad  v=U_0\pi\left(\frac{x}{L_0}-0.5\right),
\end{equation}
the disk will begin to rotate and keep its shape in the whole process.
In simulations,
the parameters are set as follows: $L_0=200, W=2, \sigma=0.01, U_0=0.02, \tau_g=1$, $\rho_A=1, \rho_B=0.5$ and $\text{Pe}=200$.
Figure~\ref{zalesak} shows the initial shape of the disk together with its final shape at $t=2T$ . As seen from this figure, all the models can give accurate results. However, both the Liang's model and model I produce a distinct deformation for the sharp corners. Furthermore, we calculate the relative errors of all models with different $\mbox{Pe}$ as shown in Fig.~\ref{fig4}.

\subsection{Circular interface in a shear flow}
\begin{figure}[!htb]
\centering
\subfloat[]{\includegraphics[width=0.25\textwidth,trim=0 20 0 20,clip]{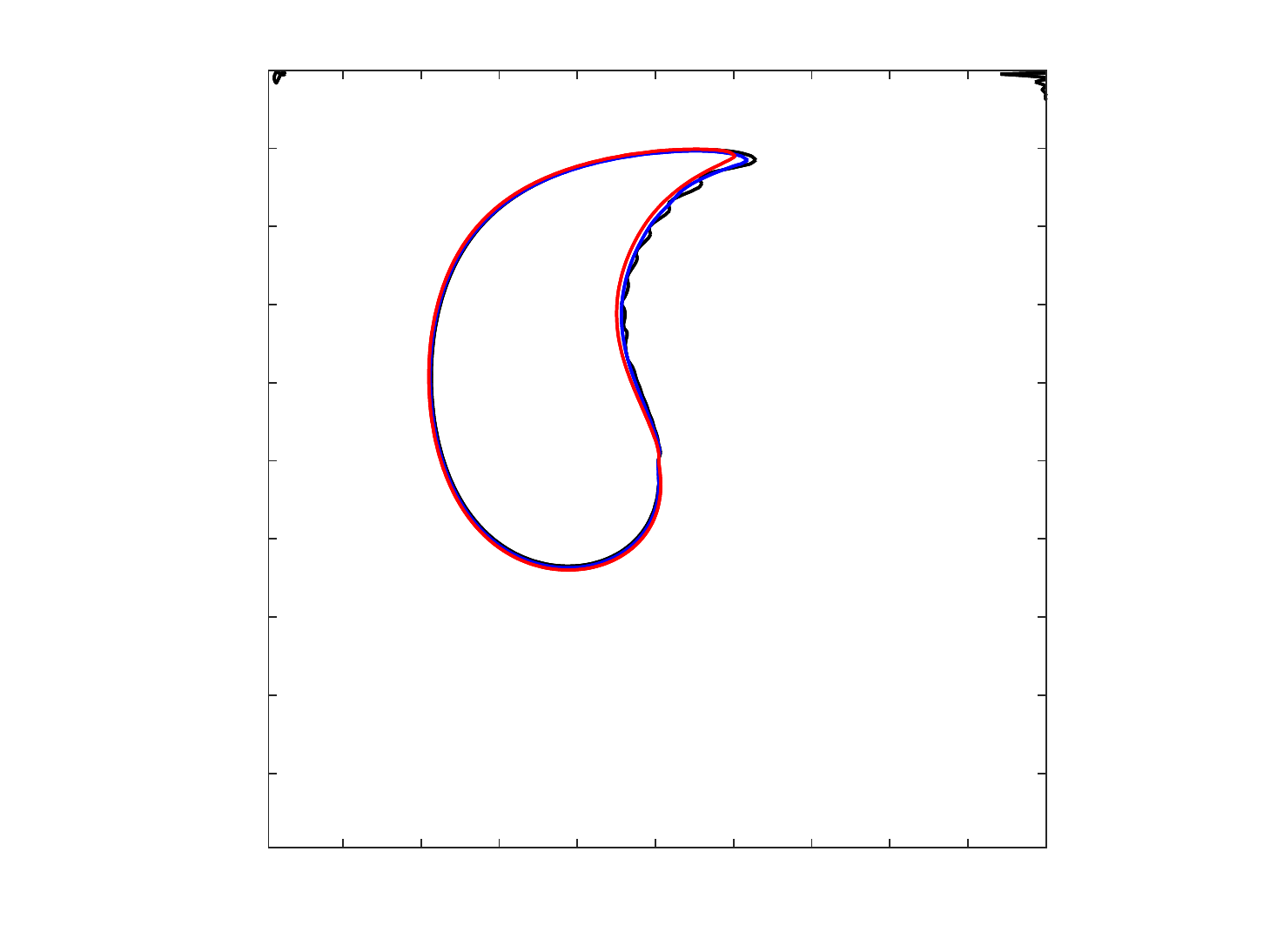}}~
\subfloat[]{\includegraphics[width=0.25\textwidth,trim=0 20 0 20,clip]{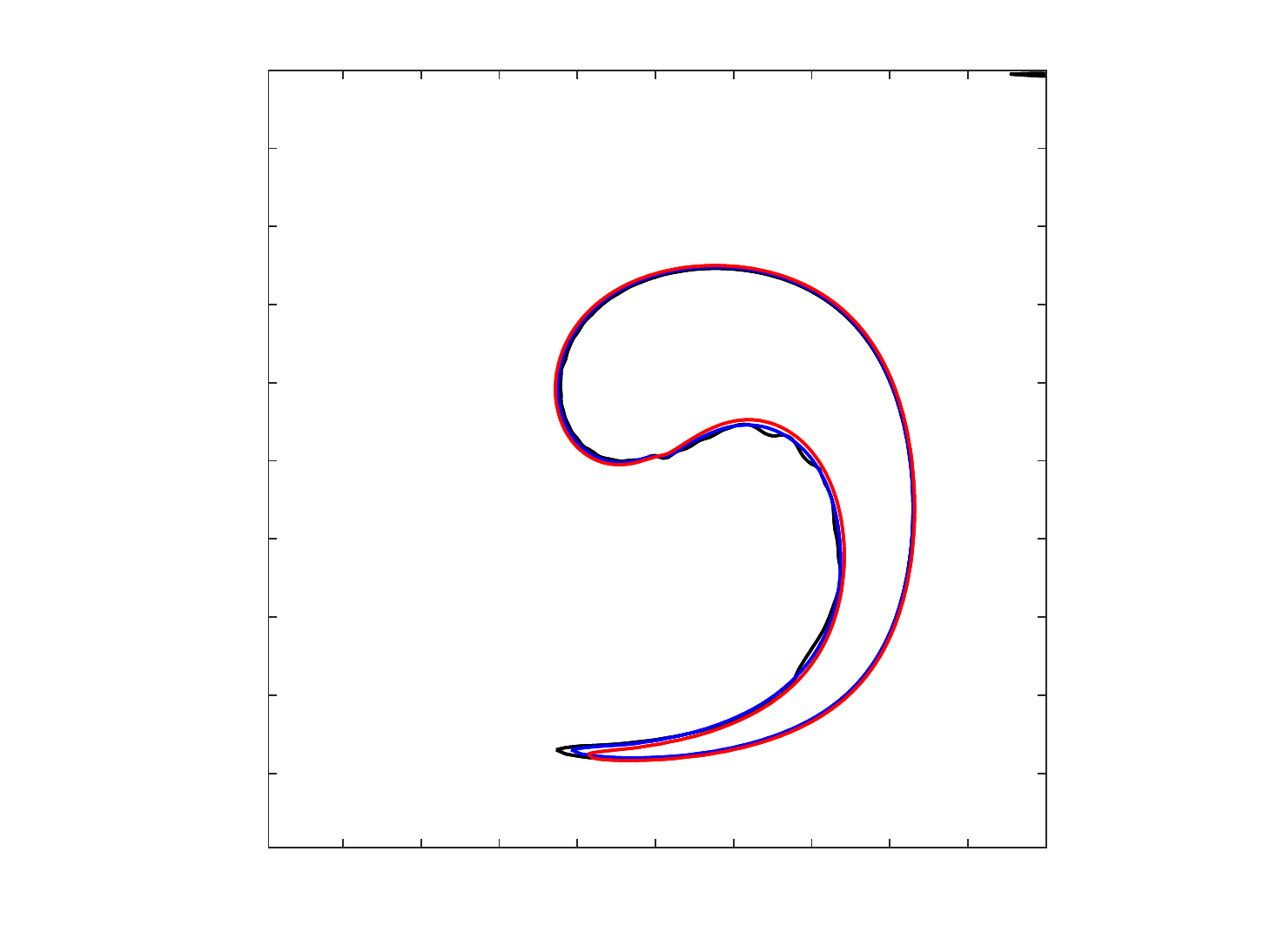}}~
\subfloat[]{\includegraphics[width=0.25\textwidth,trim=0 20 0 20,clip]{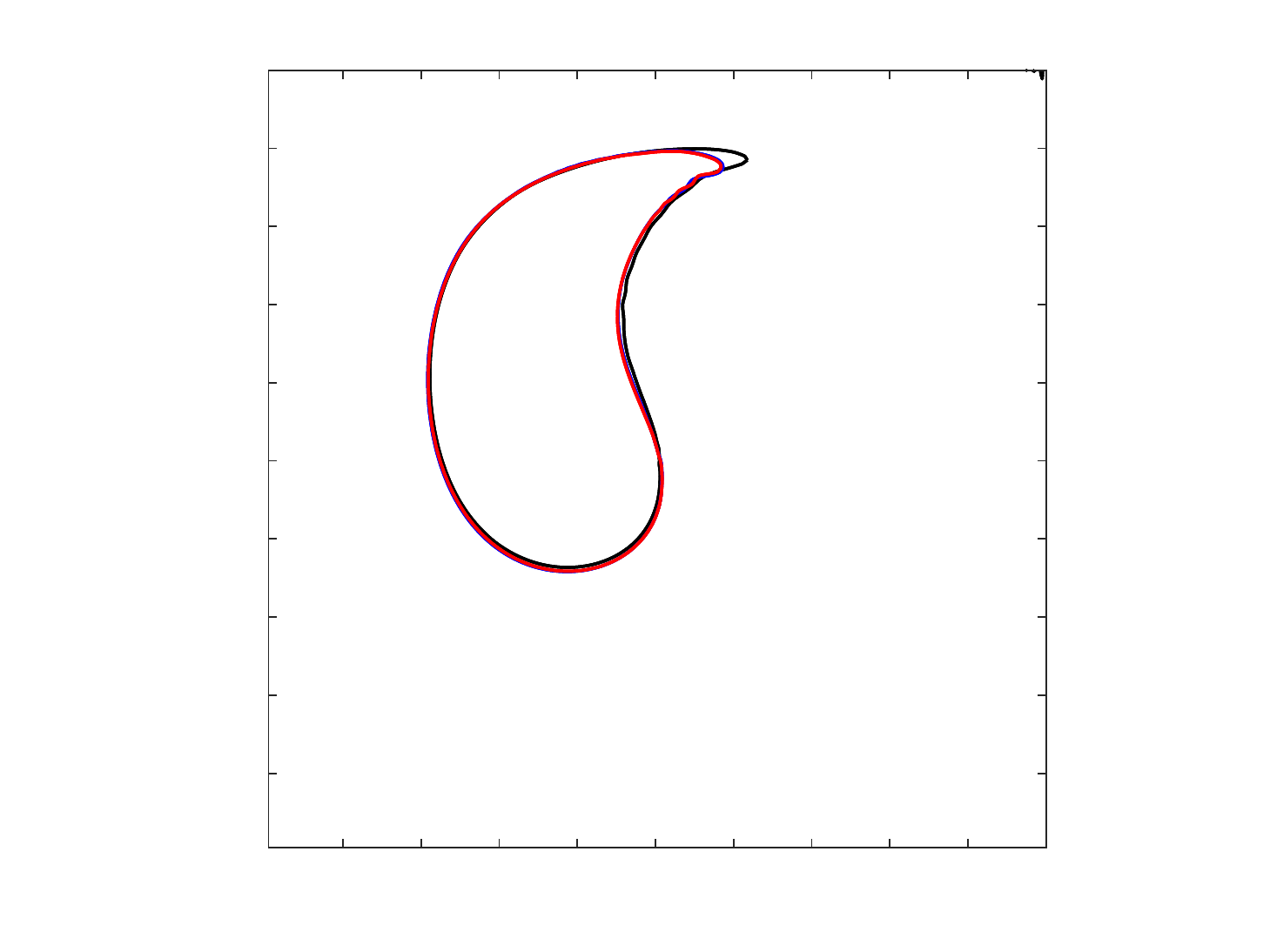}}~
\subfloat[]{\includegraphics[width=0.25\textwidth,trim=0 20 0 20,clip]{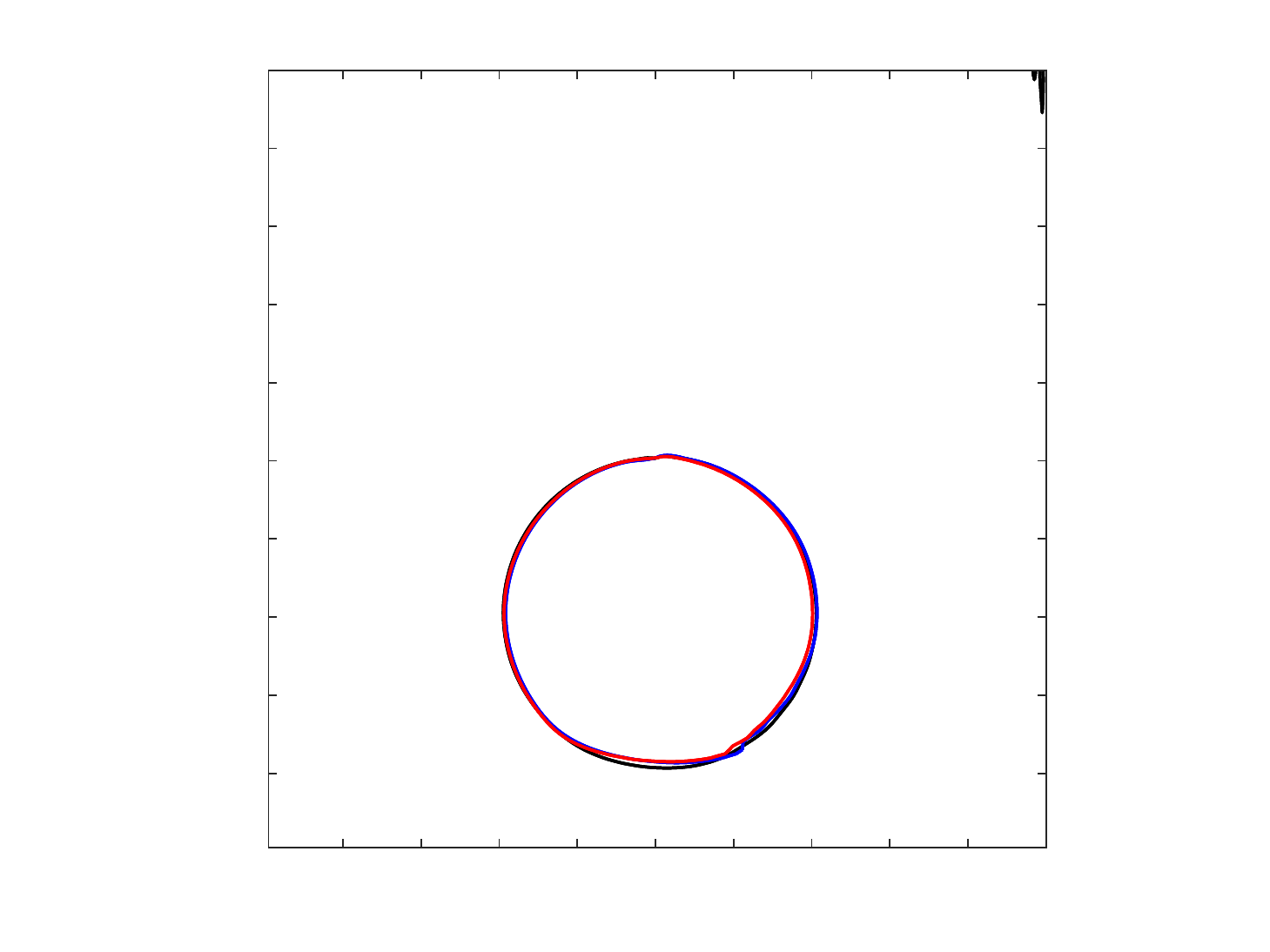}}
\caption{ The phase-field contours ($\phi=0$) of a circular interface in a shear flow at (a) 0.5T, (b) 1T, (c) 1.5T and (d) 2T.
The results of the model in \cite{liang2014phase}, model I and model II are displayed by the dash line, dashed line, dash-dotted line, respectively.}
\label{shearcompare}
\end{figure}

\begin{figure}[!htb]
  \centering
  \includegraphics[width=0.75\textwidth,trim=0 0 0 20,clip]{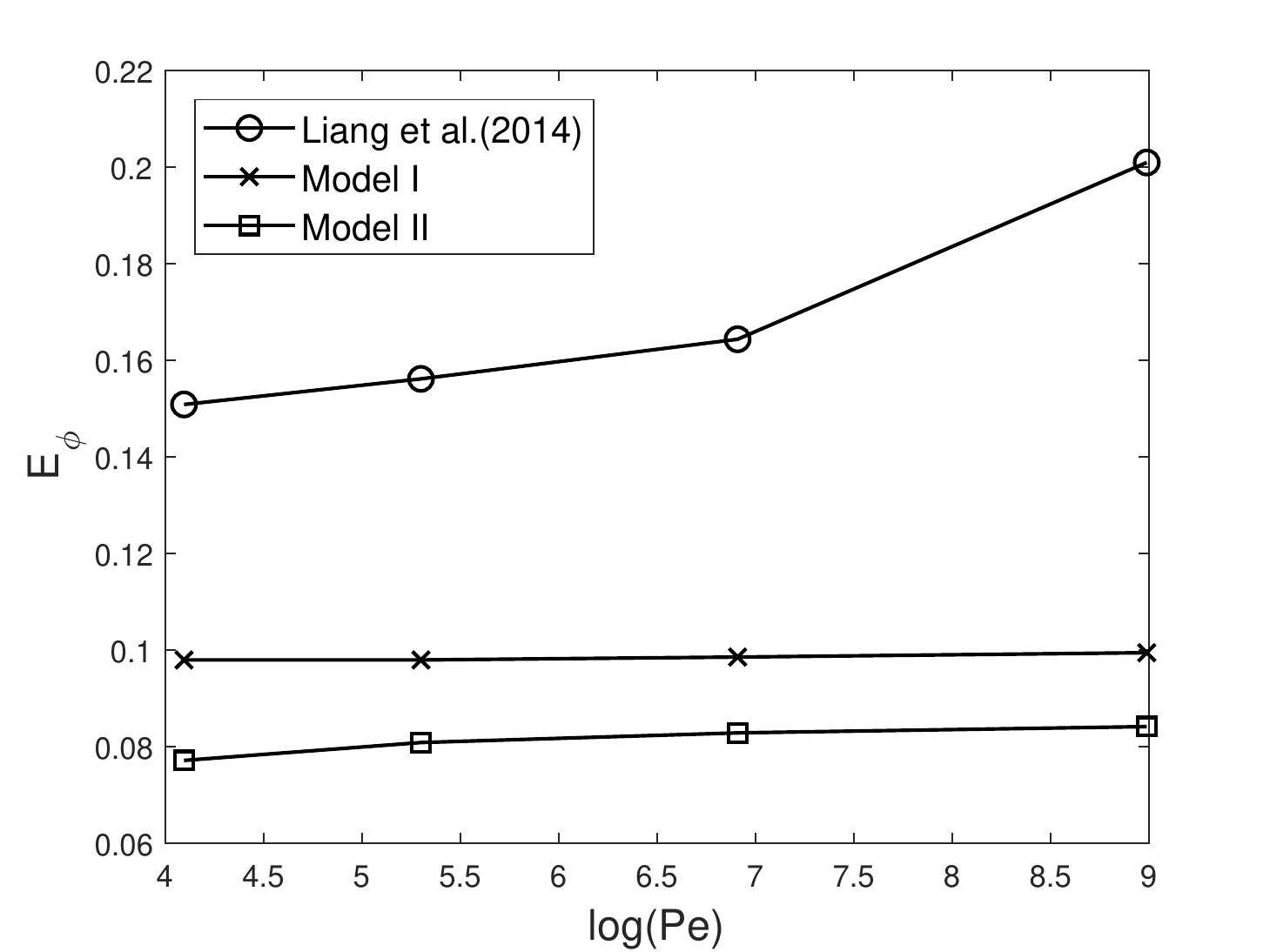}
  \caption{Relative errors of different models with $\mbox{Pe}$ numbers for shear flow. }\label{fig:case3PE}
\end{figure}

In this section, we place a circular interface into a shear flow in a domain of $L_0\times L_0$ lattice size, and the shear velocity is set to be
\begin{subequations}\label{shear}
\begin{align}
u=-U_0 \pi \cos\left[ \pi\left(\frac{x}{L_0}-0.5\right)\right] \sin\left[\pi\left(\frac{y}{L_0}-0.5\right)\right],\\
v= U_0 \pi \sin\left[ \pi\left(\frac{x}{L_0}-0.5\right)\right] \cos\left[\pi\left(\frac{y}{L_0}-0.5\right)\right].
\end{align}
\end{subequations}
Initially, the circular interface with $R=L_0/5$ is placed at $(x,y)=(100,60)$. The velocity field is reversed at $t=T$ and the interface should go back to its original position at $t=2T$. The other parameters are set as follows: $L_0=200$, $u_0=0.025$, $W=2$, $\mbox{Pe}=1000$, $\tau_g=1.0$, $\rho_A=1$, $\rho_B=0.5$, $\sigma=0.01$.  The evolution of the interface at different times is shown in Fig.~\ref{shearcompare}. It can be seen that the shape of the interface is stretched progressively into a thin filament that spirals towards the vortex center for all models.
However, the results of model III~\cite{liang2014phase} have some distortions at $t=0.5T$. At $t=1T$ and $1.5T$, the interface obtained by model II is less stretched.
Figure~\ref{fig:case3PE} shows the relative results obtained  by the proposed models and model III with different $\text{Pe}$ numbers.

\subsection{Deformation field}
\begin{figure}[!htb]
\centering
\subfloat[]{\includegraphics[width=0.25\textwidth,trim=0 20 0 20,clip]{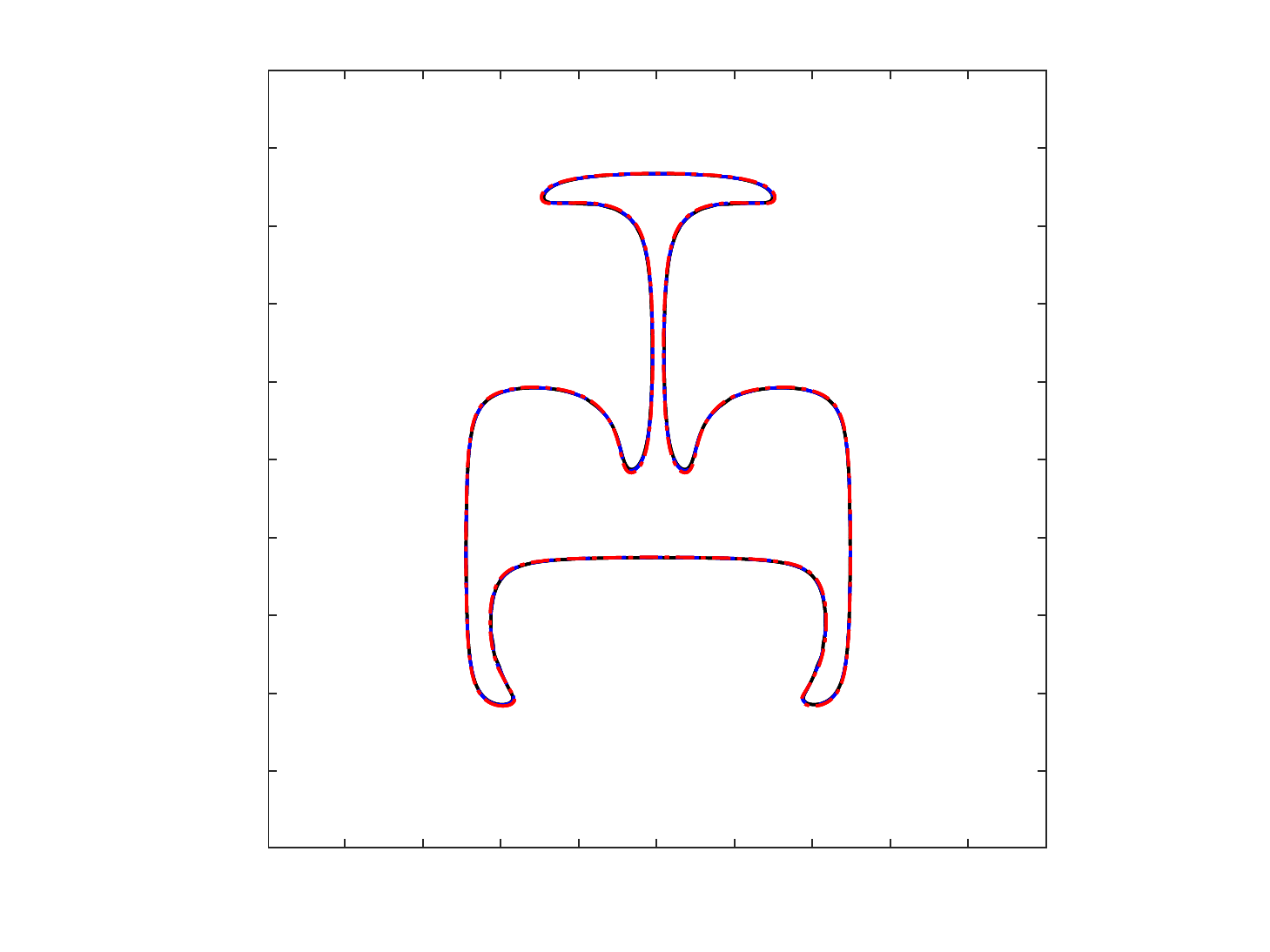}}~
\subfloat[]{\includegraphics[width=0.25\textwidth,trim=0 20 0 20,clip]{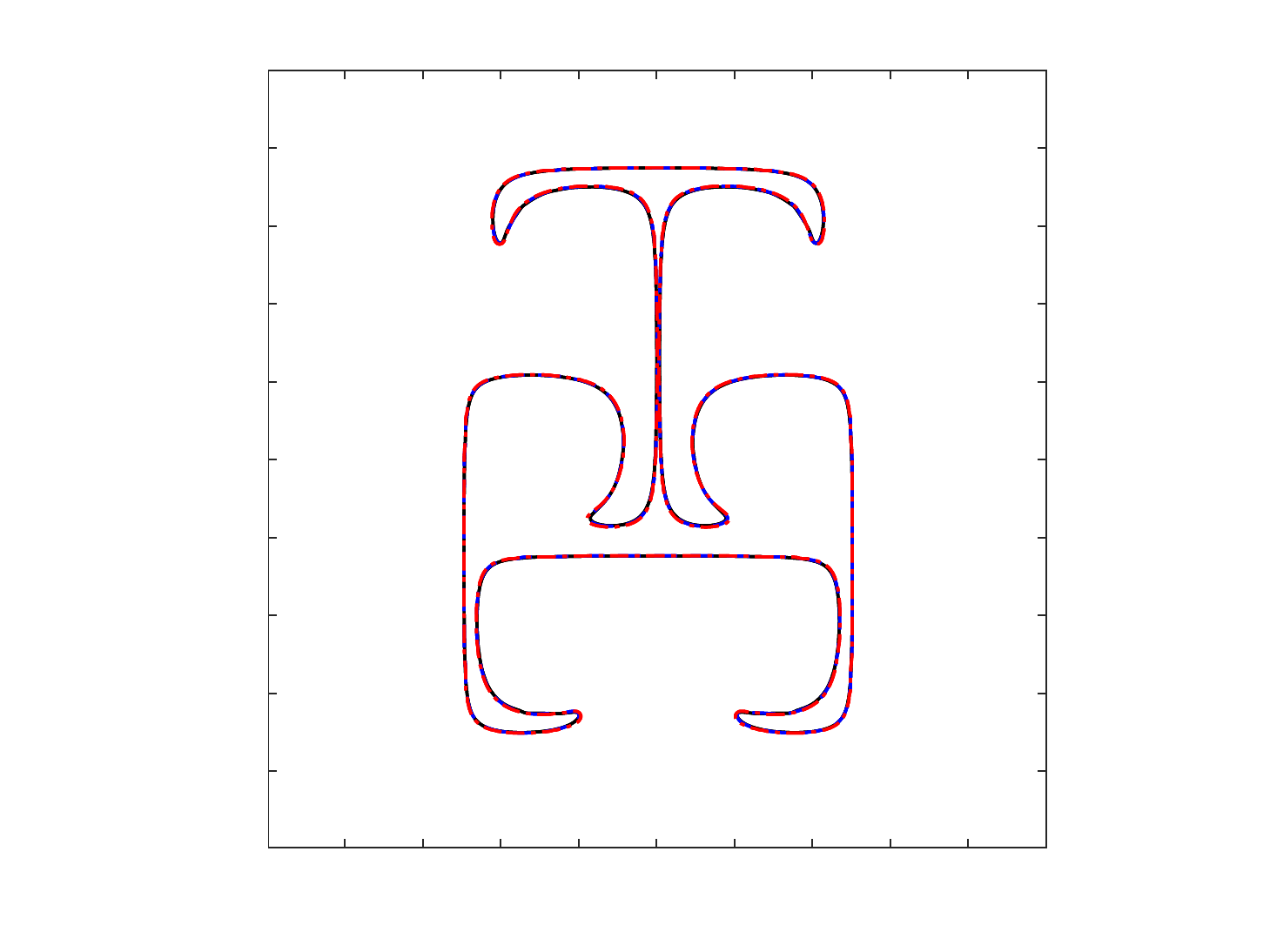}}~
\subfloat[]{\includegraphics[width=0.25\textwidth,trim=0 20 0 20,clip]{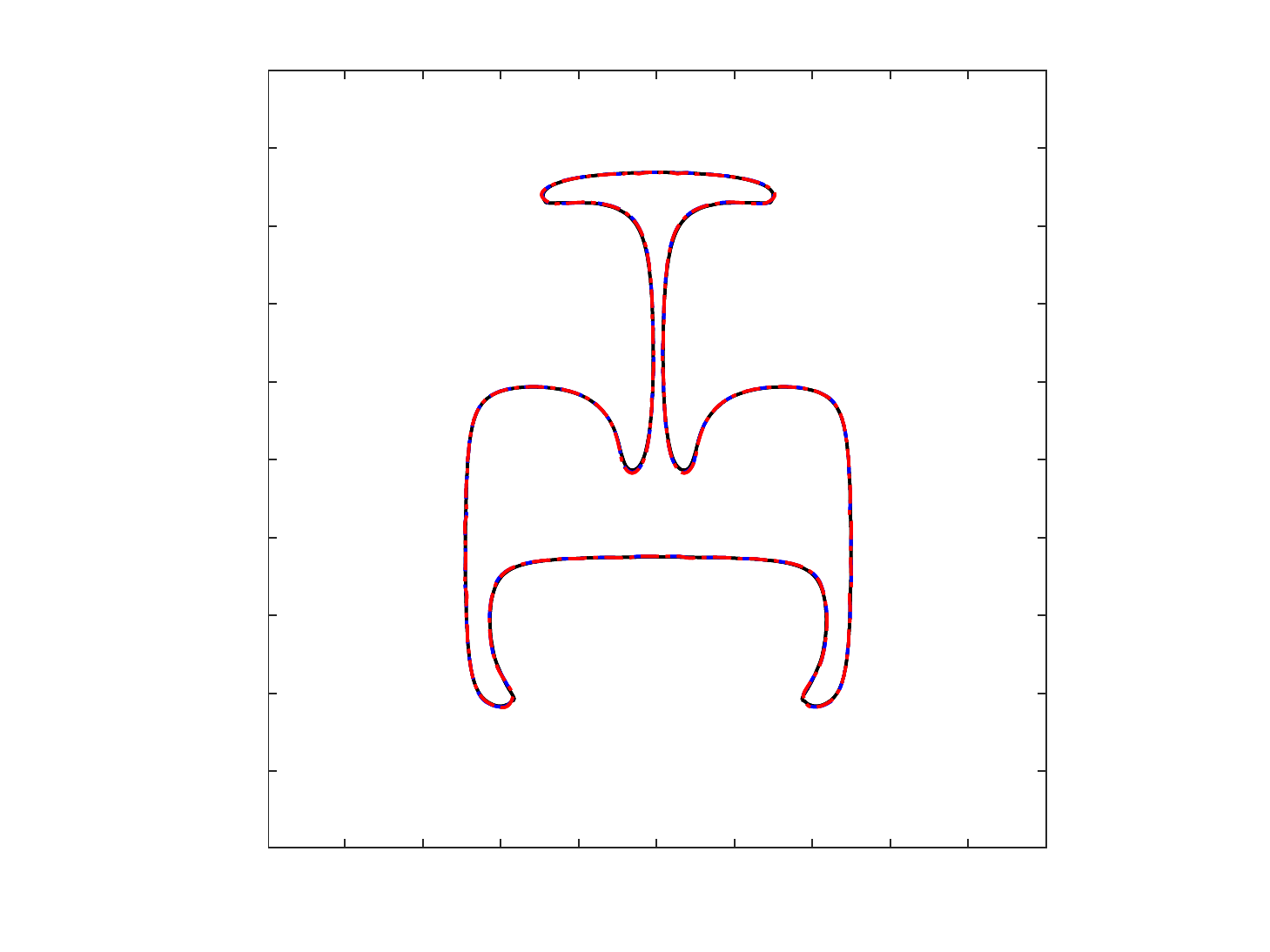}}~
\subfloat[]{\includegraphics[width=0.25\textwidth,trim=0 20 0 20,clip]{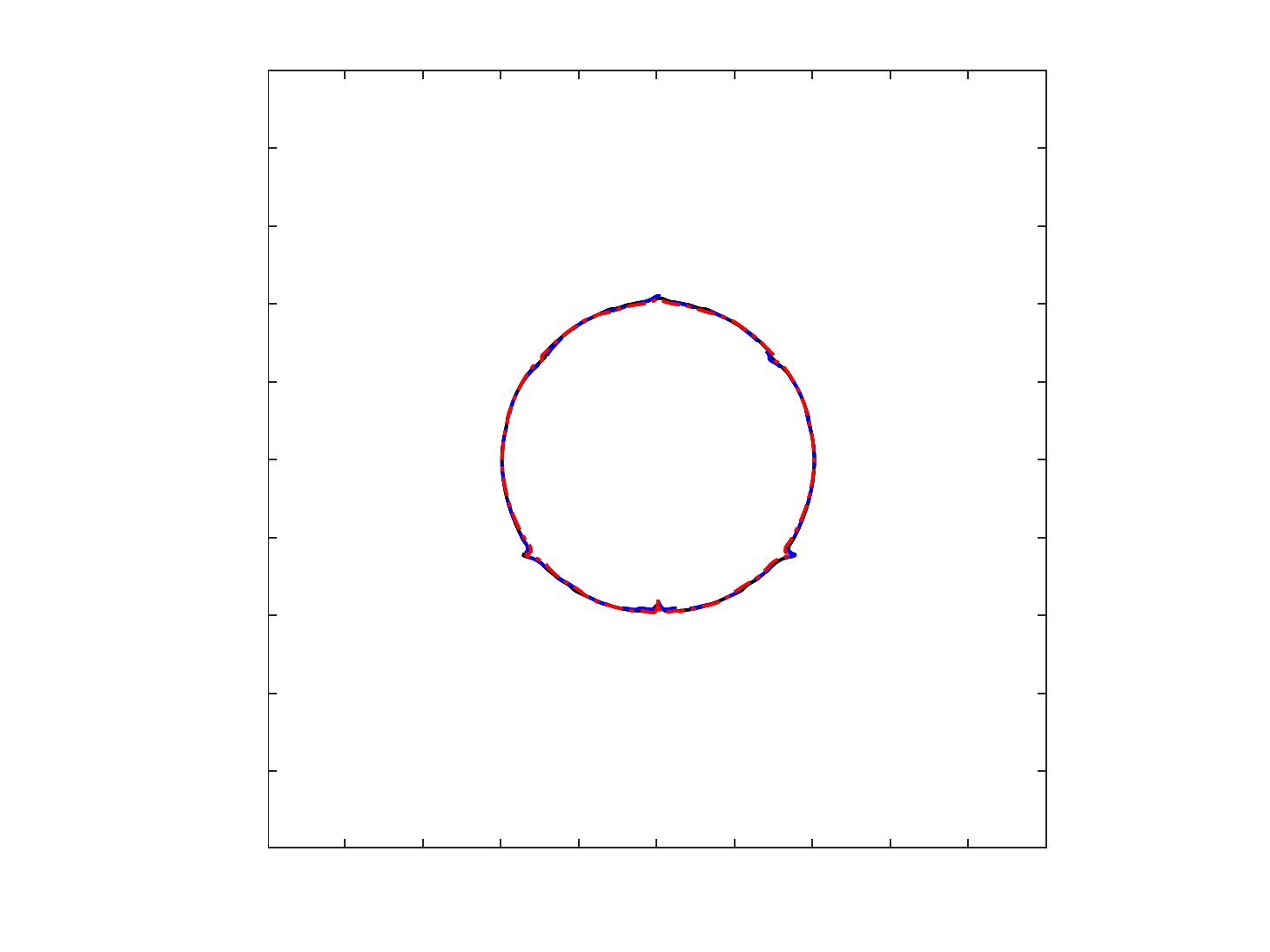}}
\caption{ The phase-field contours ($\phi=0$) at (a) 0.5T, (b) 1T, (c) 1.5T and (d) 2T.
The results of the model in \cite{liang2014phase}, model I and model II are displayed by the solid line, dashed line, dash-dotted line, respectively.}
\label{case4comparephi0}
\end{figure}

\begin{figure}[!htb]
\centering
    \subfloat[]{
    \includegraphics[width=0.2\textwidth,trim=0 25 0 23,clip]{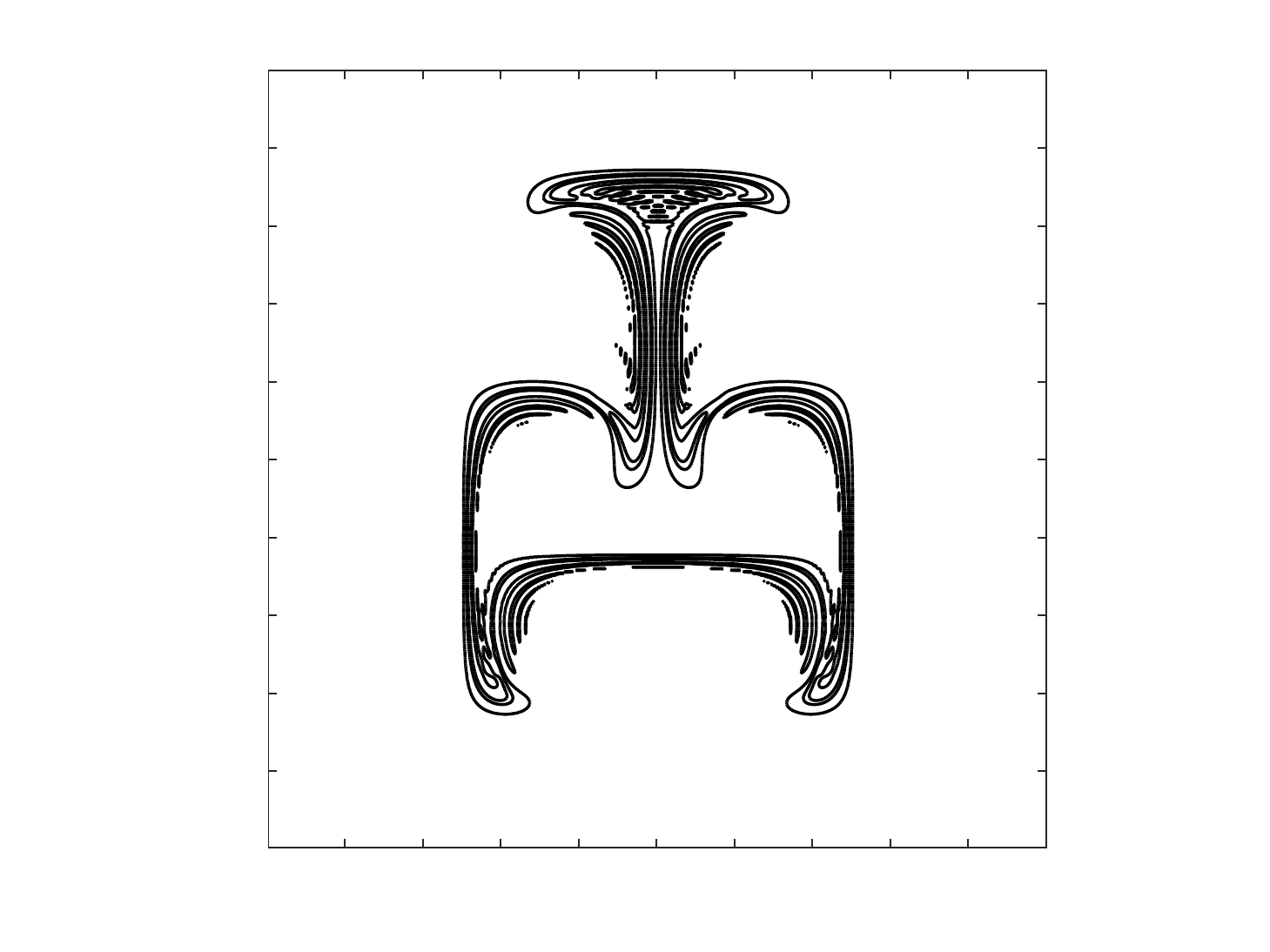}~
    \includegraphics[width=0.2\textwidth,trim=0 25 0 23,clip]{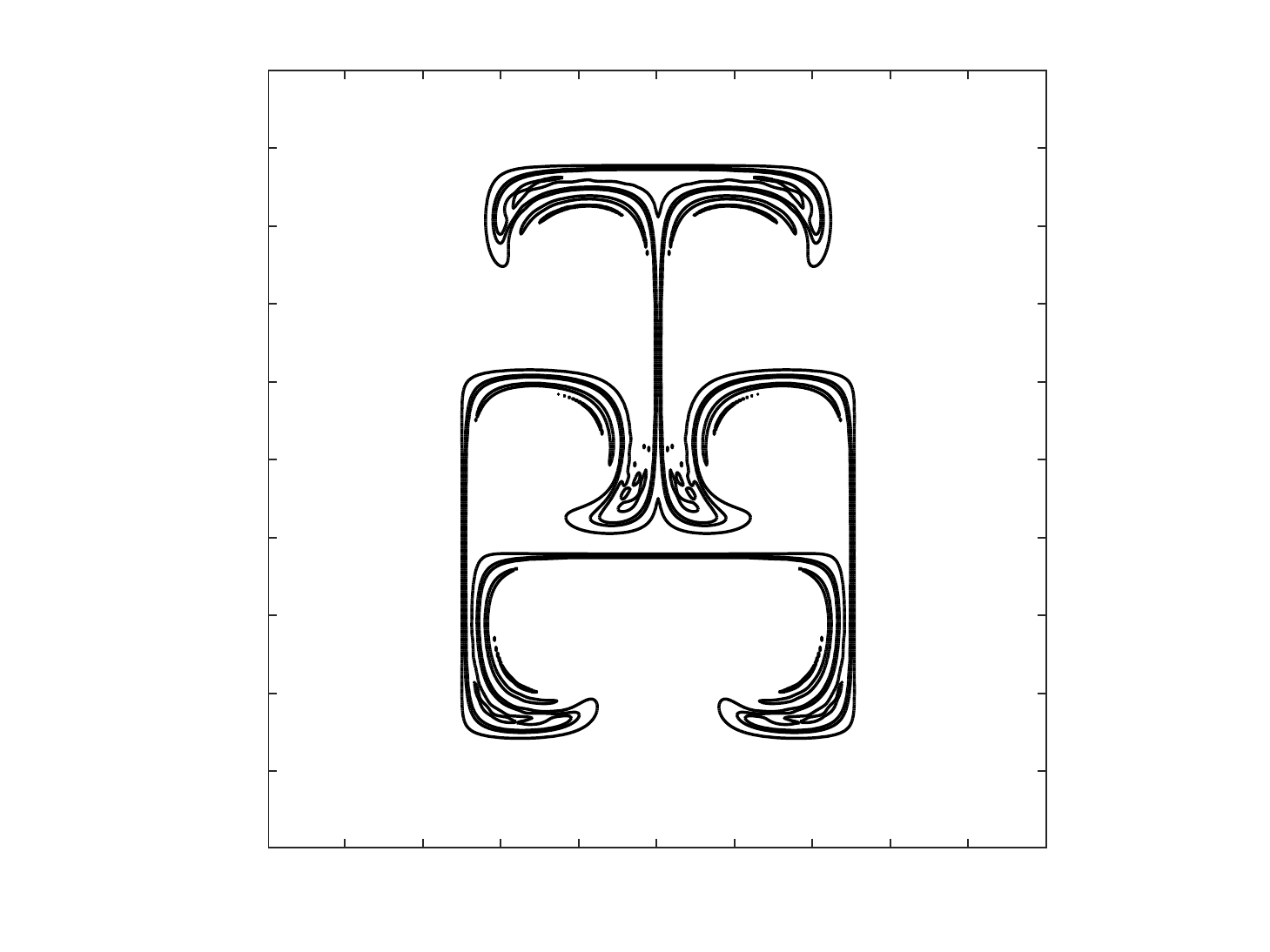}~
    \includegraphics[width=0.2\textwidth,trim=0 25 0 23,clip]{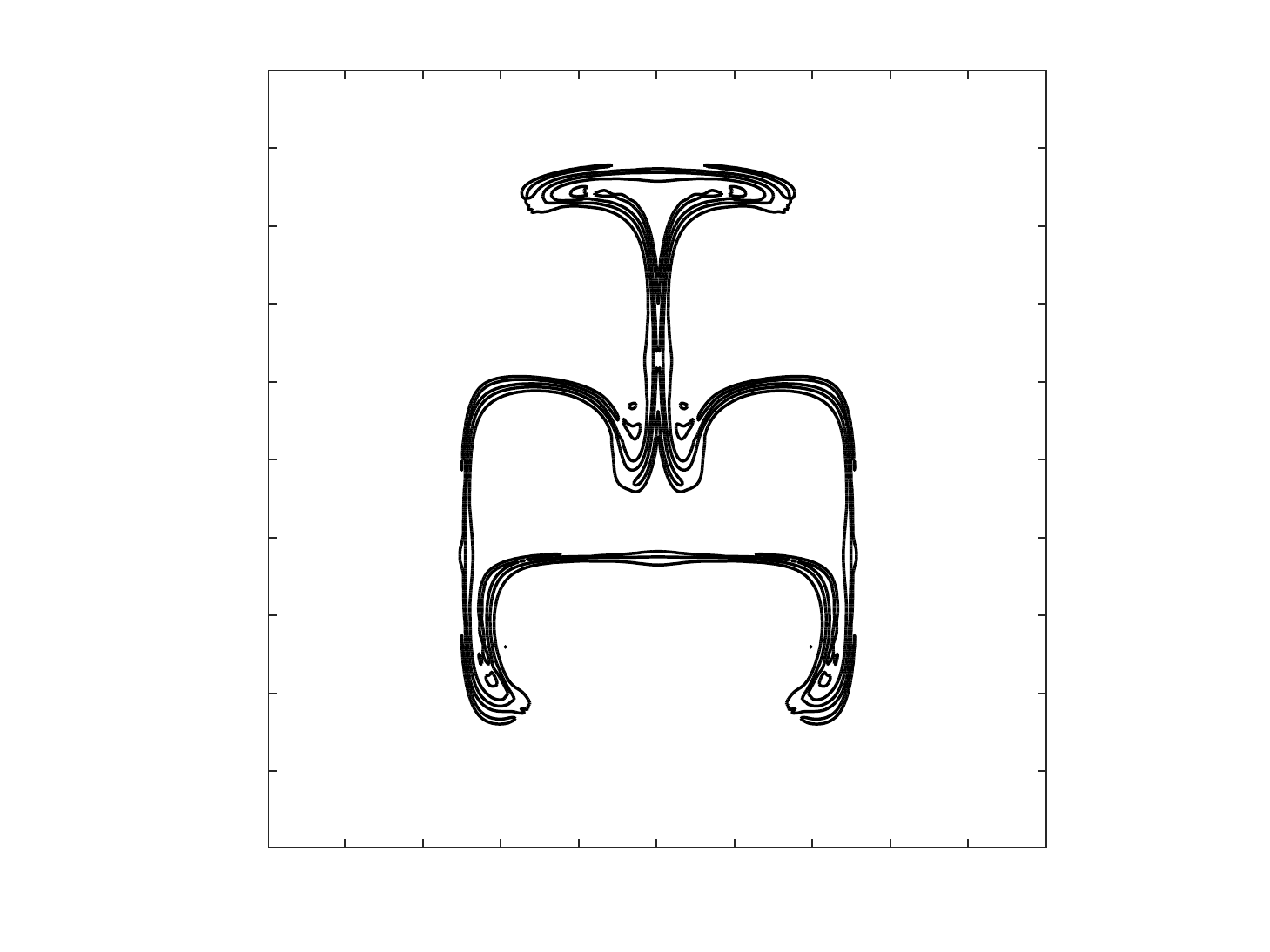}~
    \includegraphics[width=0.2\textwidth,trim=0 25 0 23,clip]{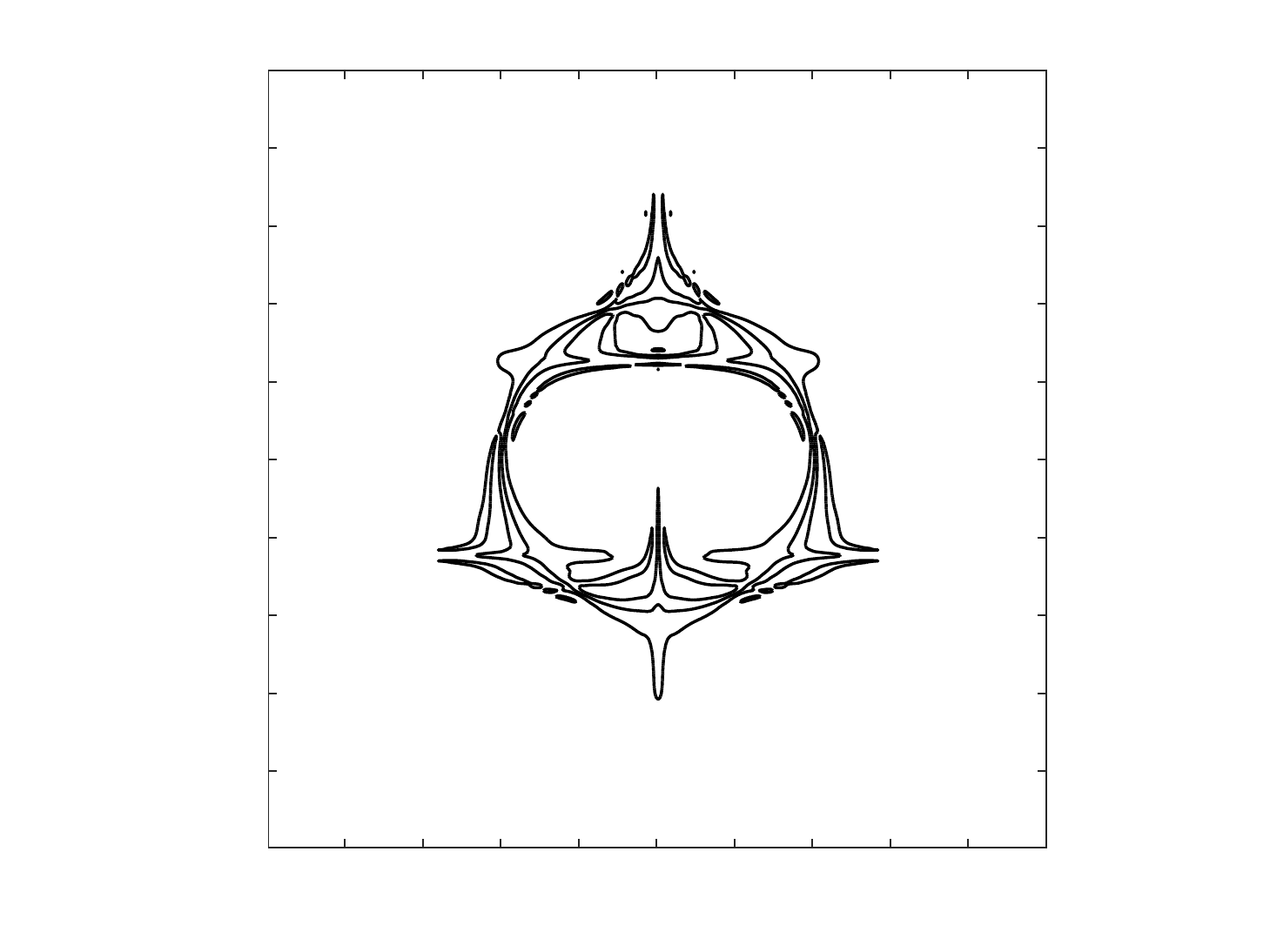}}\\
    \subfloat[]{
    \includegraphics[width=0.2\textwidth,trim=0 25 0 23,clip]{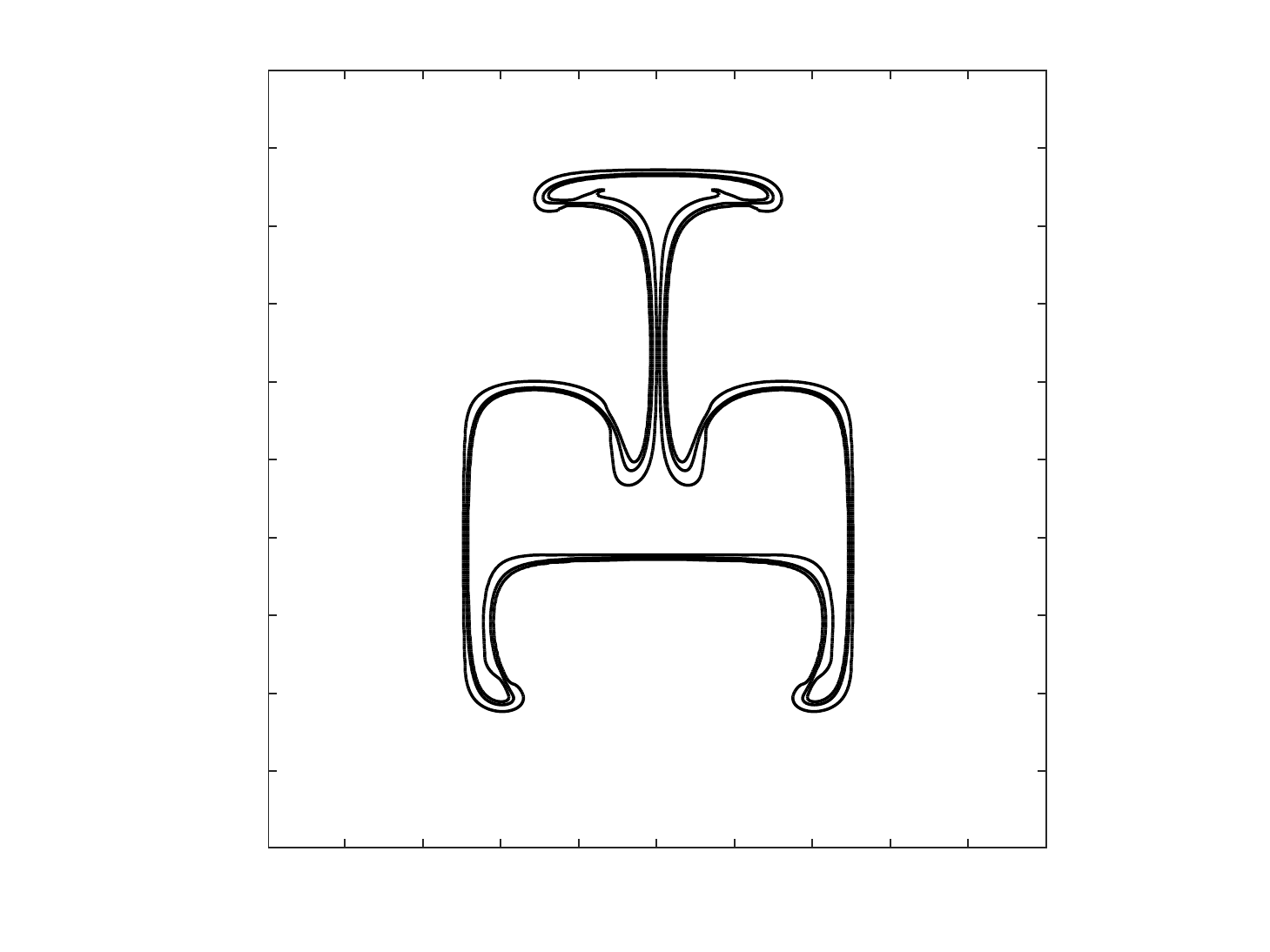}~
    \includegraphics[width=0.2\textwidth,trim=0 25 0 23,clip]{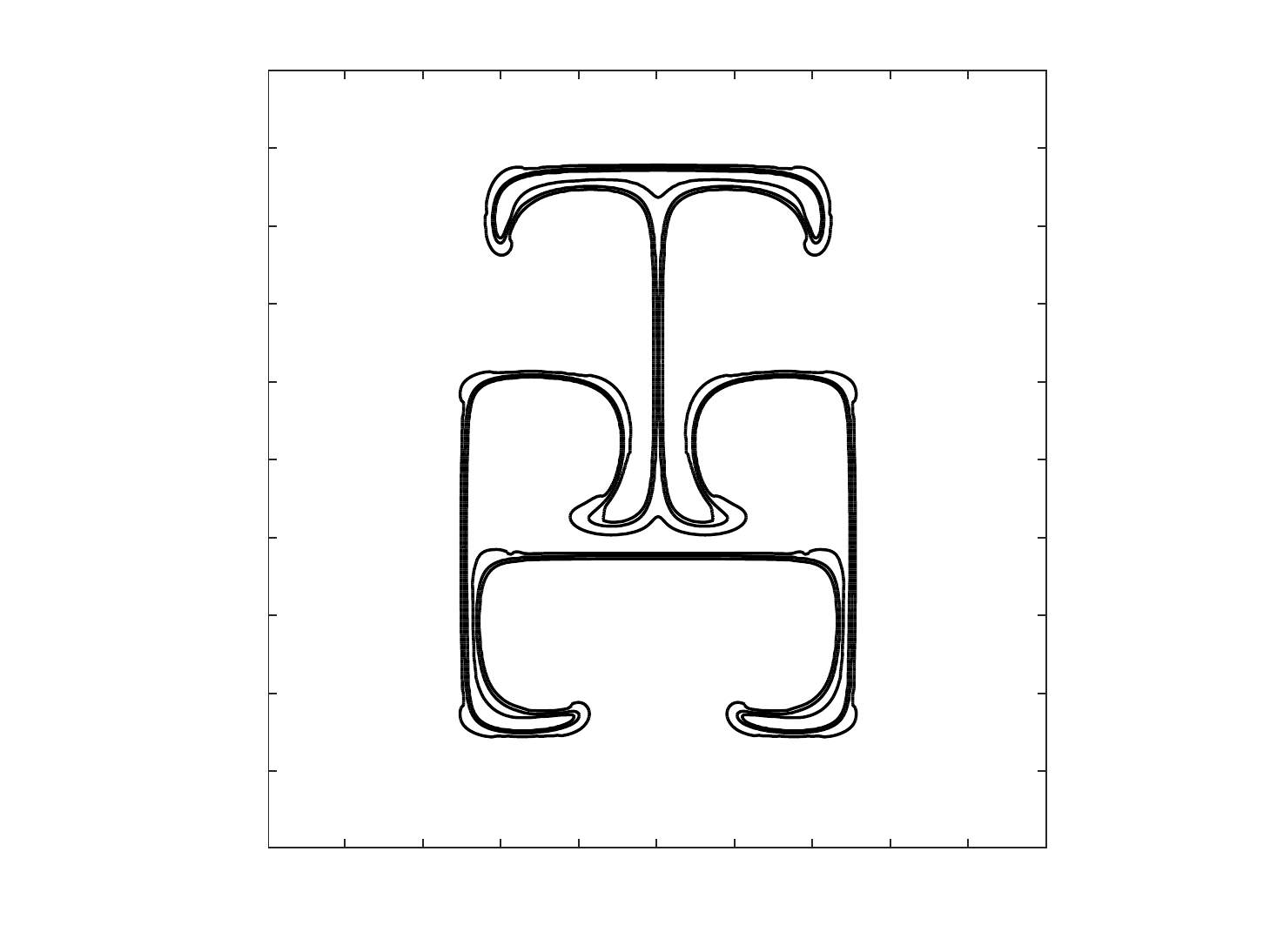}~
    \includegraphics[width=0.2\textwidth,trim=0 25 0 23,clip]{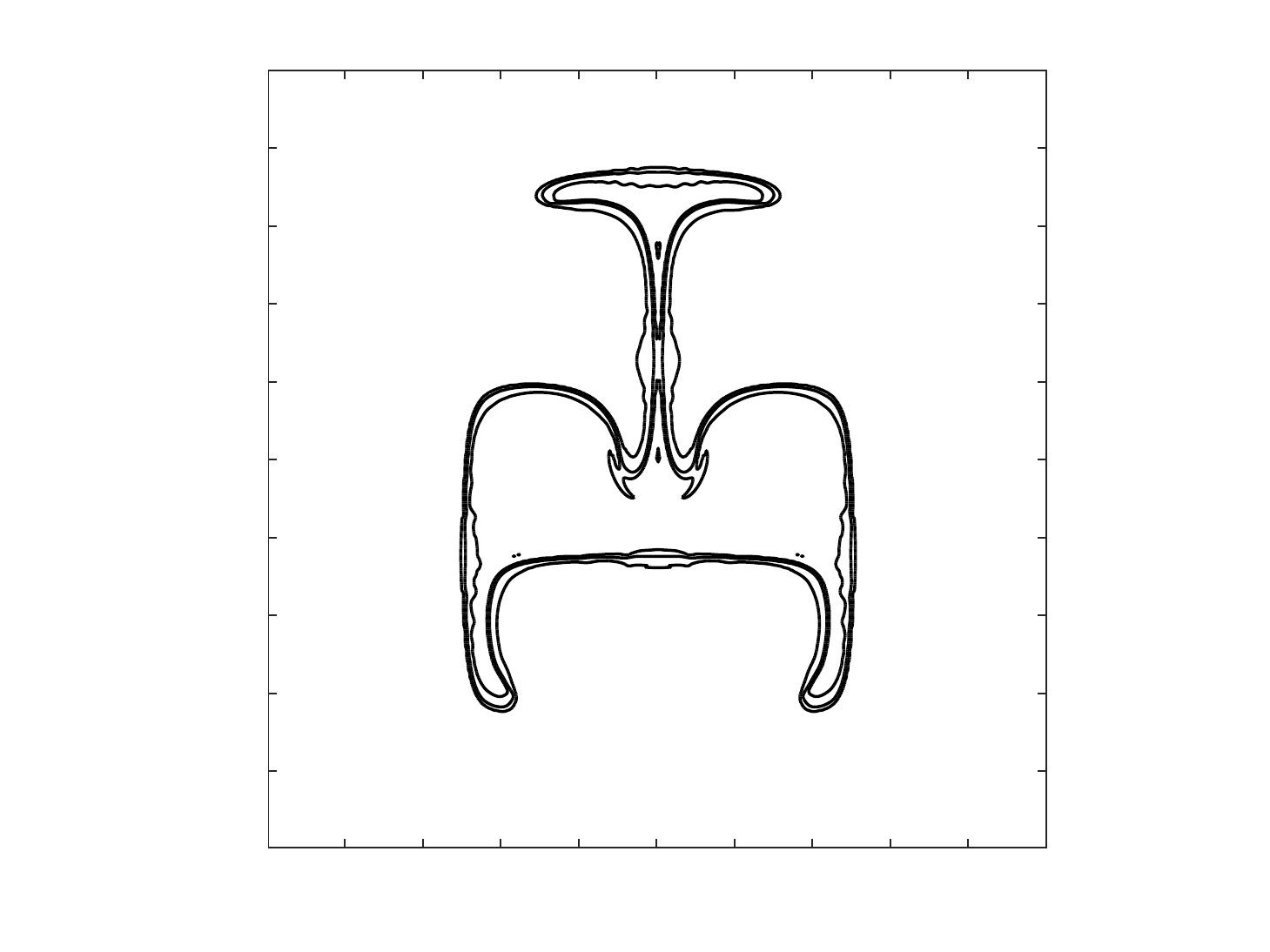}~
    \includegraphics[width=0.2\textwidth,trim=0 25 0 23,clip]{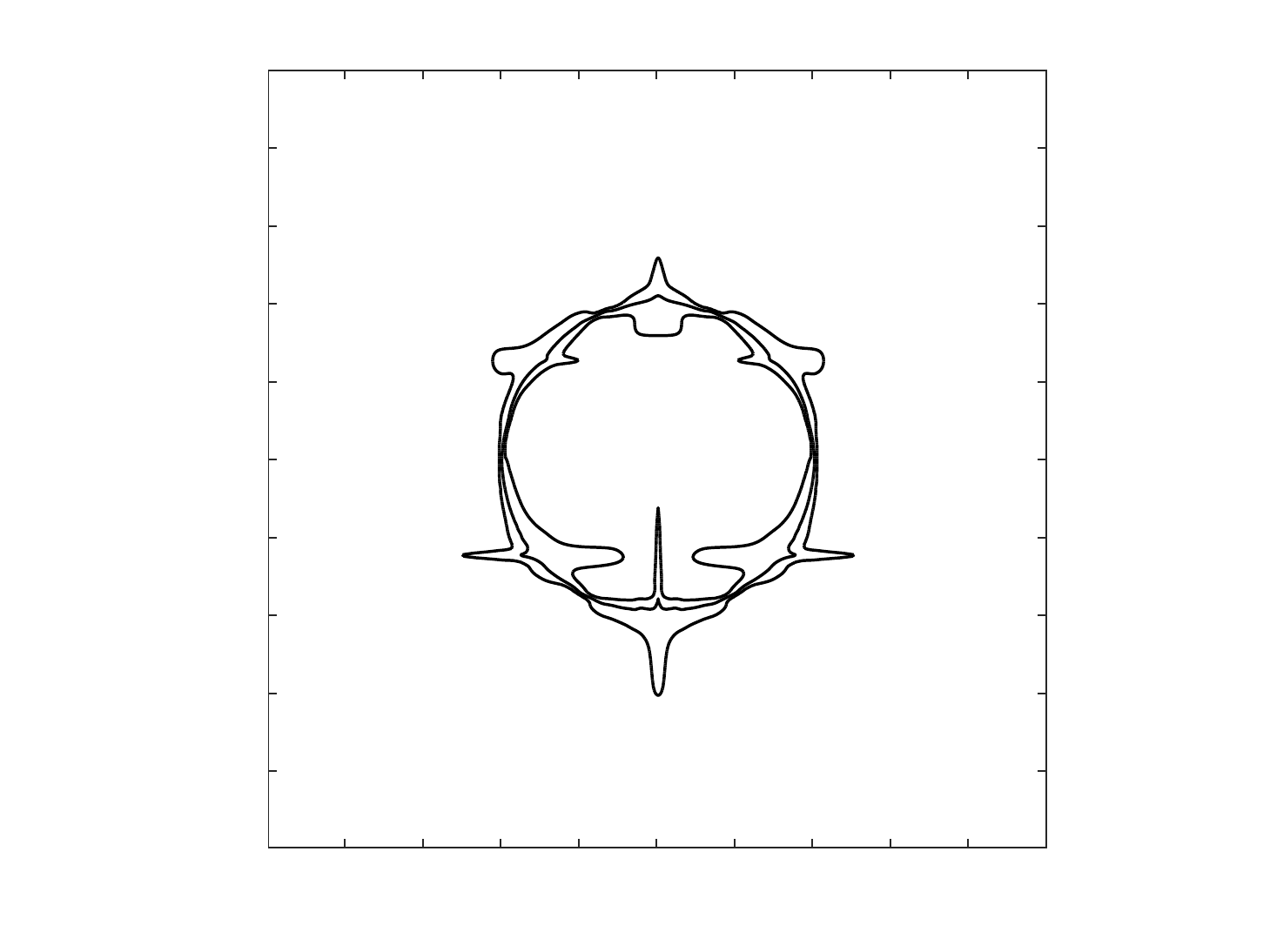}}\\
    \subfloat[]{
    \includegraphics[width=0.2\textwidth,trim=0 25 0 23,clip]{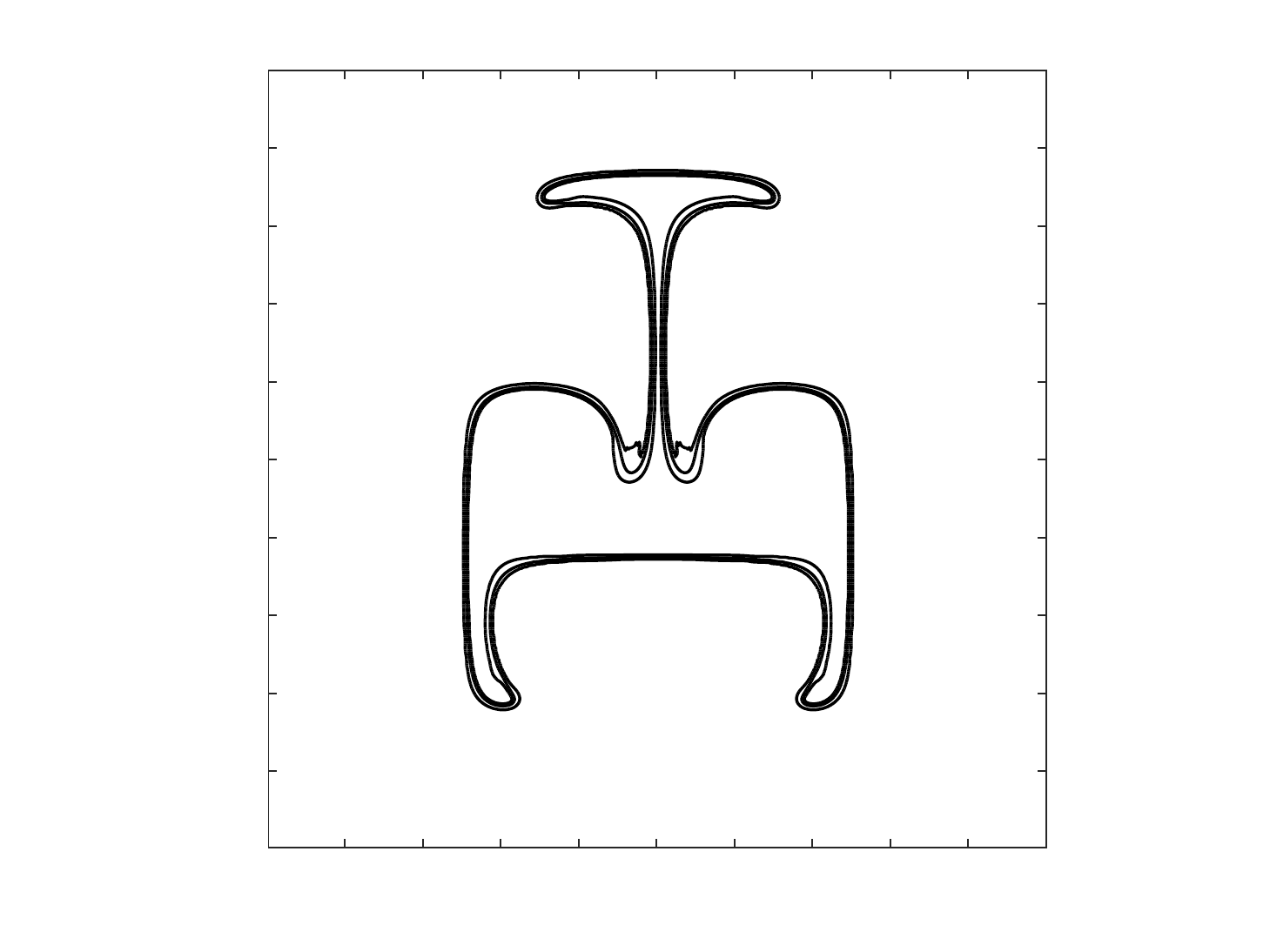}~
    \includegraphics[width=0.2\textwidth,trim=0 25 0 23,clip]{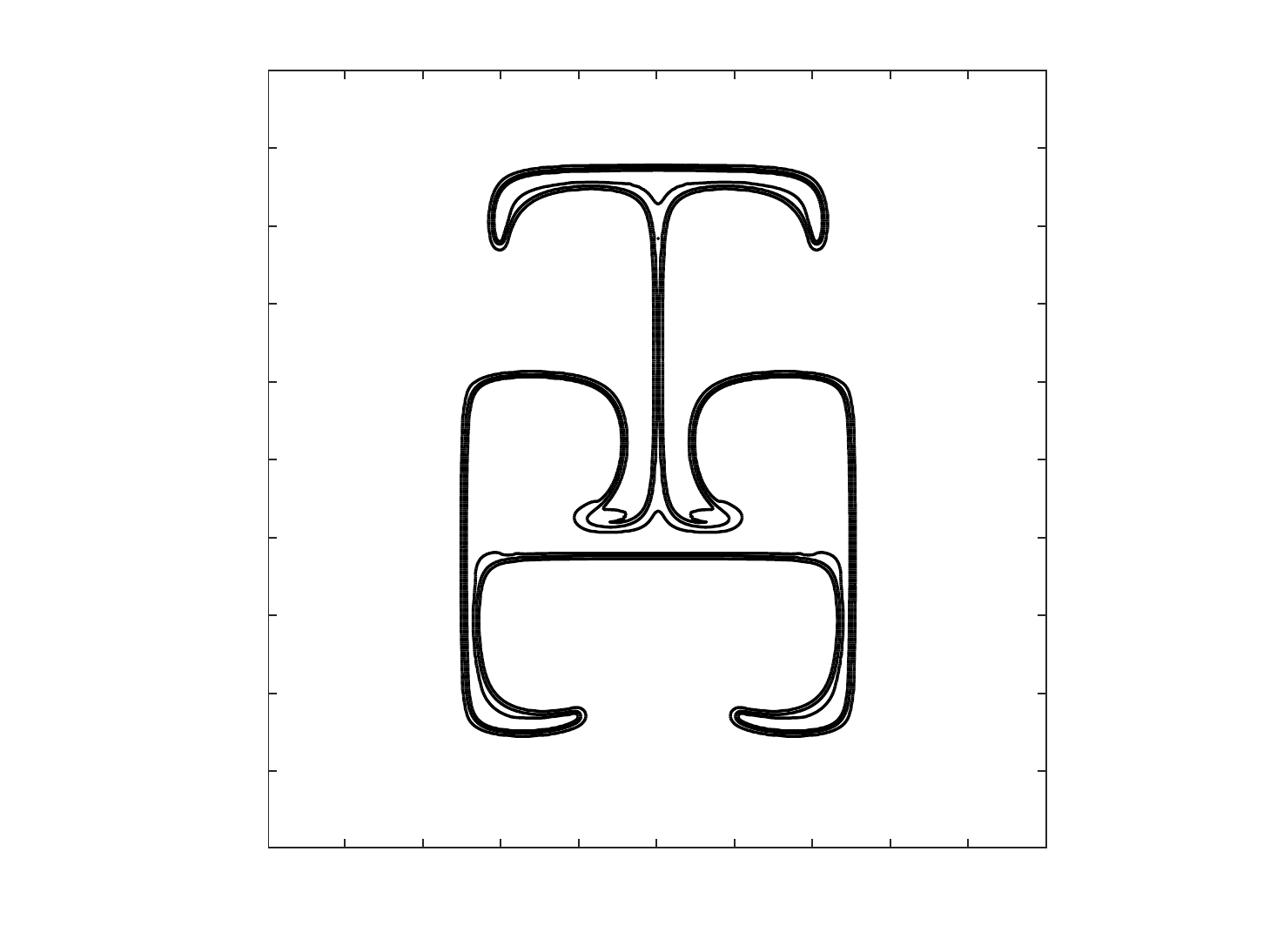}~
    \includegraphics[width=0.2\textwidth,trim=0 25 0 23,clip]{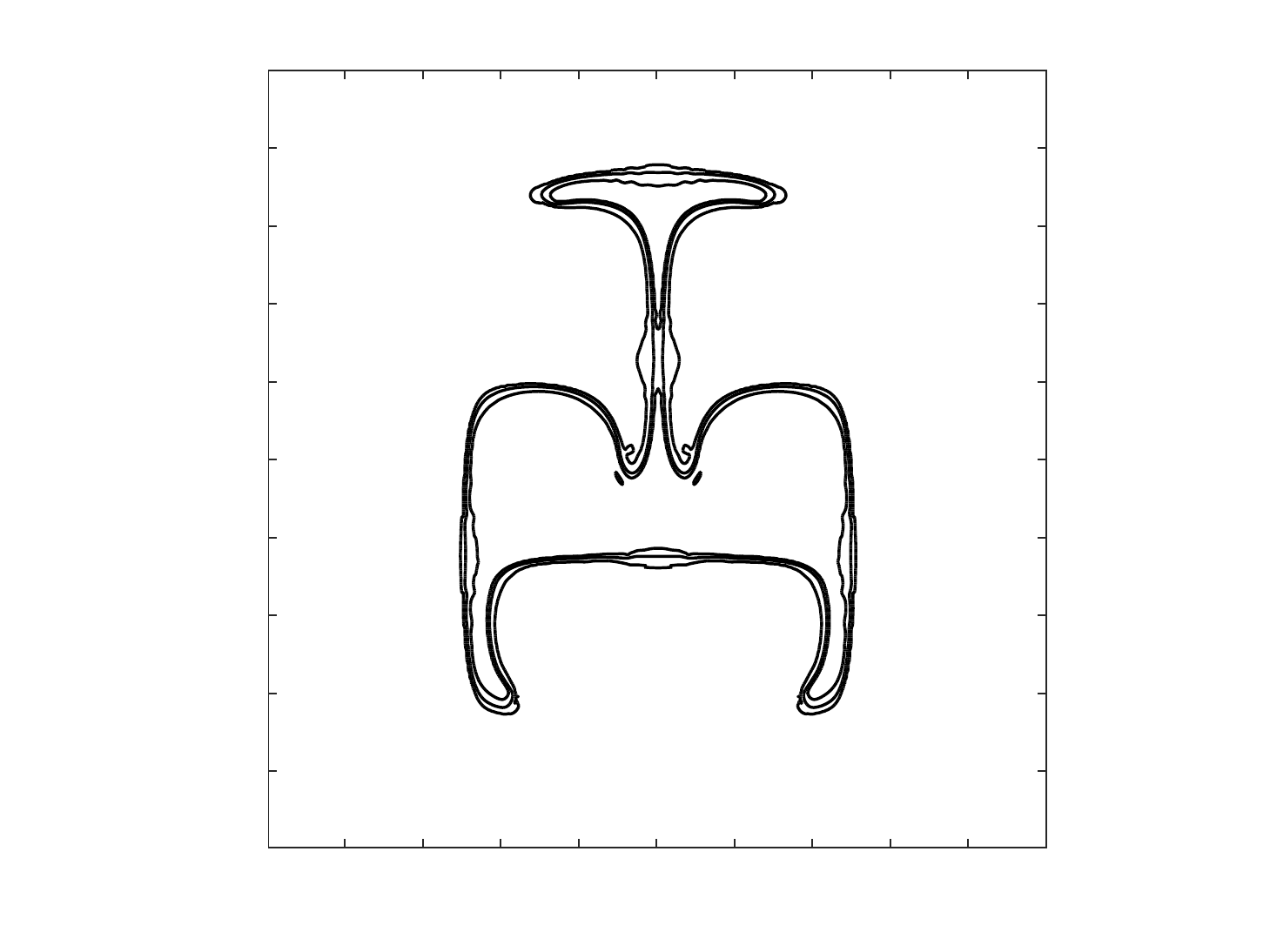}~
    \includegraphics[width=0.2\textwidth,trim=0 25 0 23,clip]{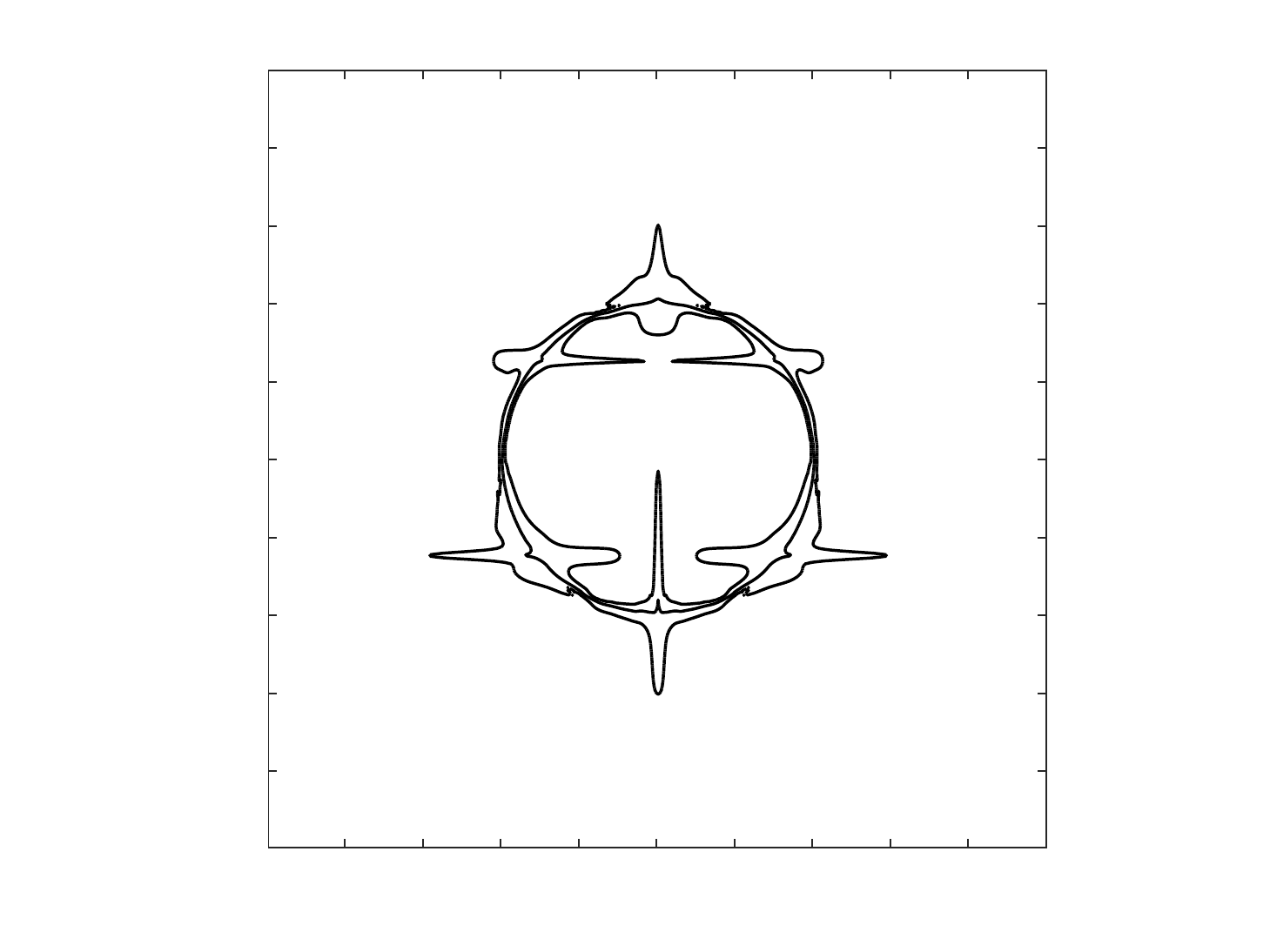}}\\
\caption{ Snapshots for a circular interface in the deformation field  with (a) the model in \cite{liang2014phase}, (b) model I, and (c) model II. From left to right:   $t=T/4, T/2, 3T/4, T$. The contours at $\phi=-0.95, 0 ,0.95$ are displayed.}\label{deformationEvo}
\end{figure}

\begin{figure}[!htb]
  \centering
  \includegraphics[width=0.75\textwidth,trim=0 1 0 20,clip]{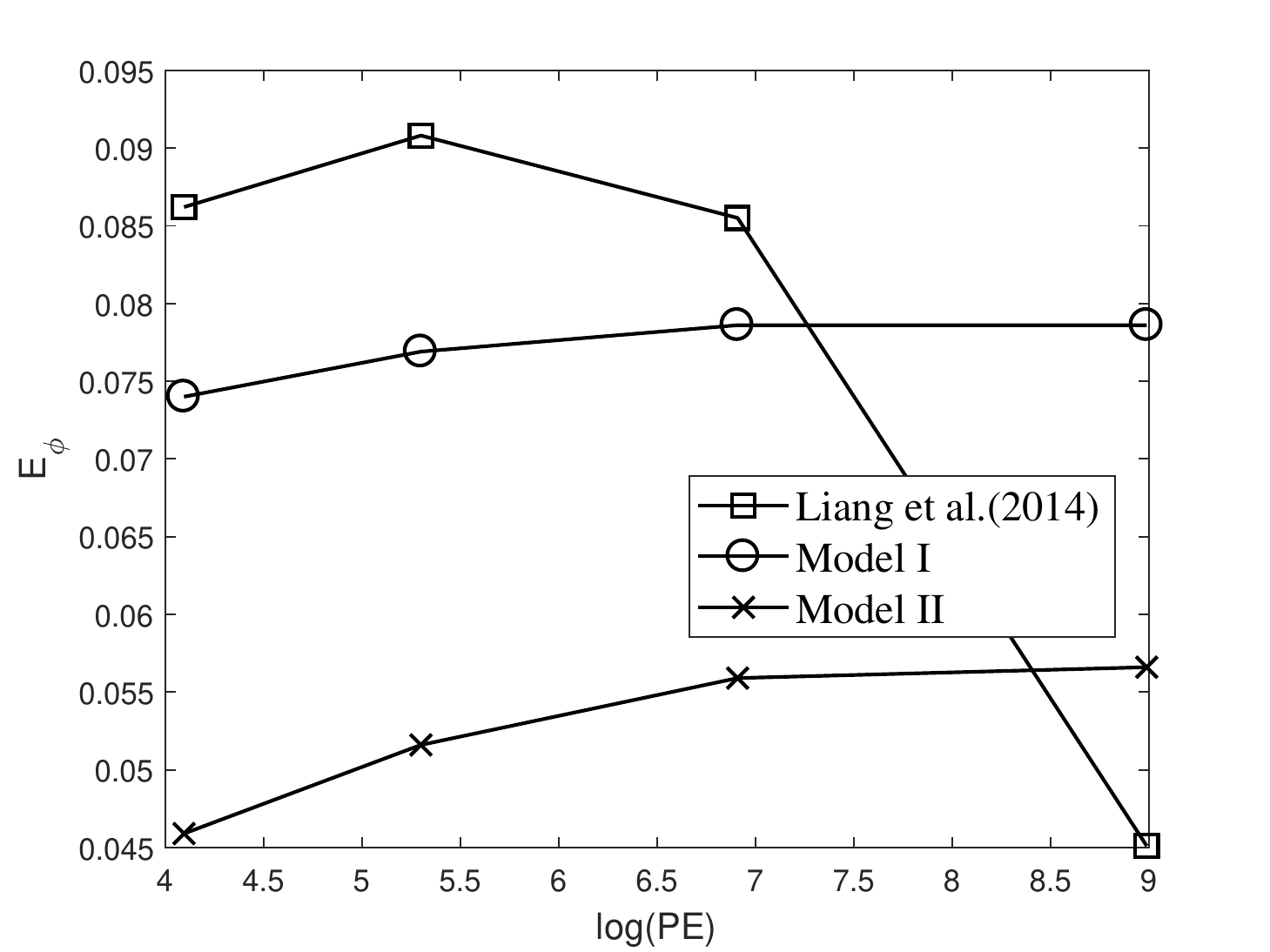}
  \caption{Relative errors of different models with $\mbox{Pe}$ numbers for the deformation field.}\label{deformationErr}
\end{figure}

In order to test the capacity of the present models, a more complicated problem with deformation field is implemented. For this problem, a time-dependent and strongly nonlinear velocity field is given by
\begin{subequations}\label{deformation}
\begin{align}
u=-U_0 \sin\left[ n\pi\left(\frac{x}{L_0}+0.5\right)\right] \sin\left[n\pi\left(\frac{y}{L_0}+0.5\right)\right]\cos\frac{\pi t}{T_0},\\
v=-U_0 \cos\left[ n\pi\left(\frac{x}{L_0}+0.5\right)\right] \cos\left[n\pi\left(\frac{y}{L_0}+0.5\right)\right]\cos\frac{\pi t}{T_0},
\end{align}
\end{subequations}
where $T_0=5T/4$, $n$ is the number of vortices and fixed to be $4$.
Initially, a circular interface with radius $R=L_0/5$ is placed in the middle of the computational domain $L_0\times L_0$. In simulations, the parameters are set as the following: $L_0=500$, $\sigma=0.01$, $U_0=0.025$ and $W=2$. The evolution of the interface patterns for all models is shown in Fig.~\ref{case4comparephi0}.
It can be seen that the shapes of the interface captured by  all models are very similar. To further comparison, the contours of $\phi=-0.95, 0, 0.95$ are plotted in Fig.~\ref{deformationEvo}. we can observe that the variation of the interface thickness simulated by the model II are the smoothest among three models.
Figure~\ref{deformationErr} presents the relative errors for the proposed models and the model in \cite{liang2014phase} with different values of $\mbox{Pe}$.

\subsection{Rayleigh-Taylor instability}
\begin{figure}[!htb]
\centering
\subfloat[]{
    \includegraphics[width=0.2\textwidth,trim=145 20 145 23,clip]{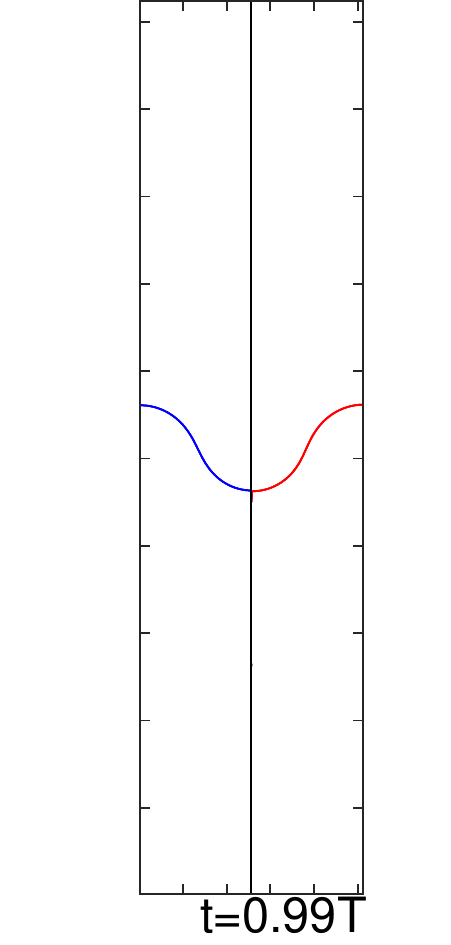}~
    \includegraphics[width=0.2\textwidth,trim=145 20 145 23,clip]{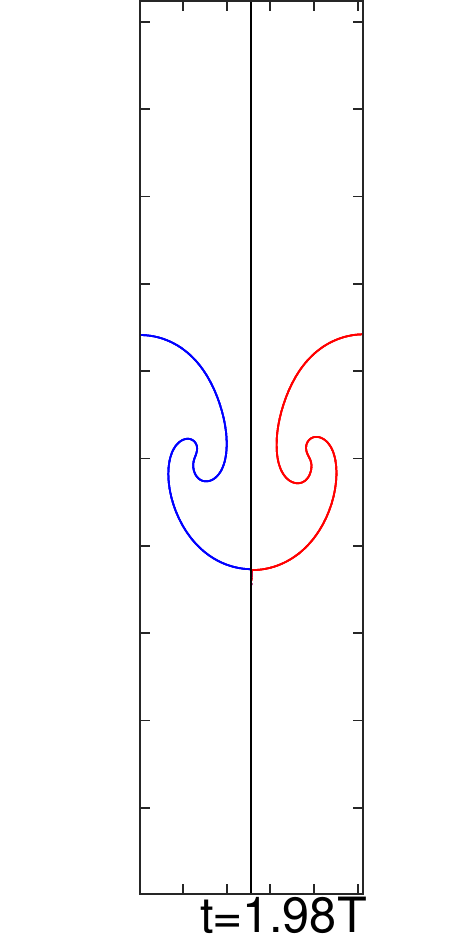}~
    \includegraphics[width=0.2\textwidth,trim=145 20 145 23,clip]{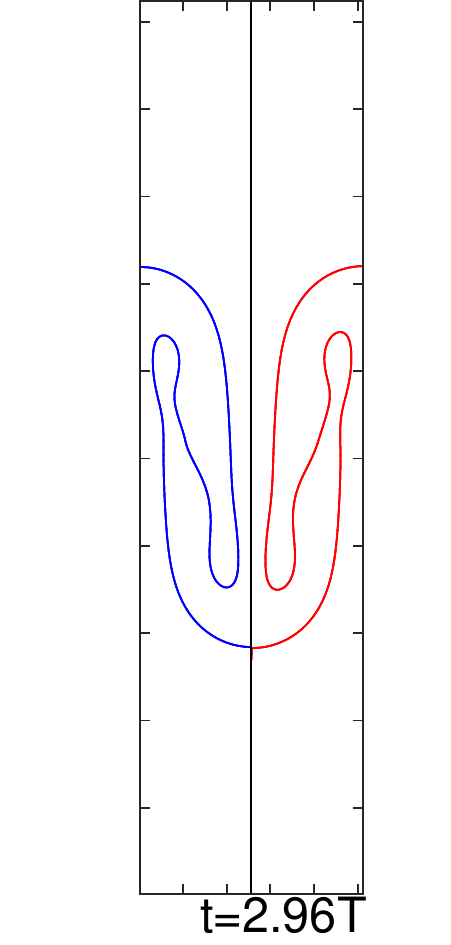}~
    \includegraphics[width=0.2\textwidth,trim=145 20 145 23,clip]{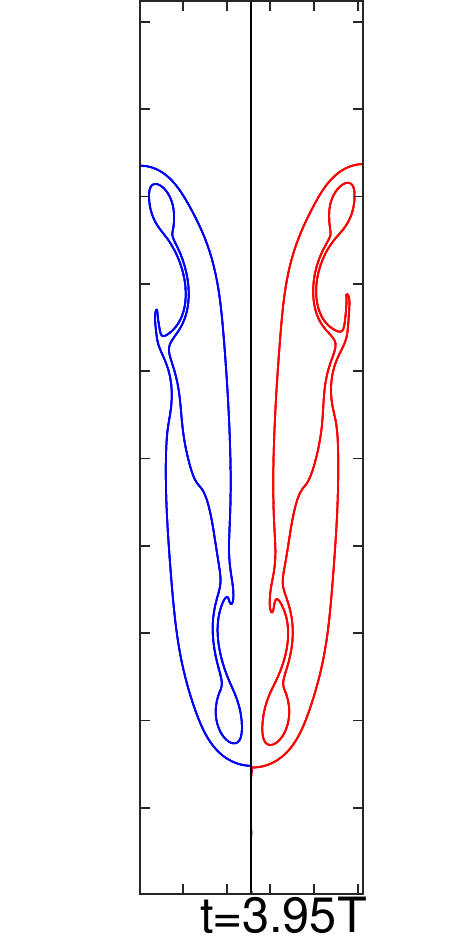}~
    \includegraphics[width=0.2\textwidth,trim=145 20 145 23,clip]{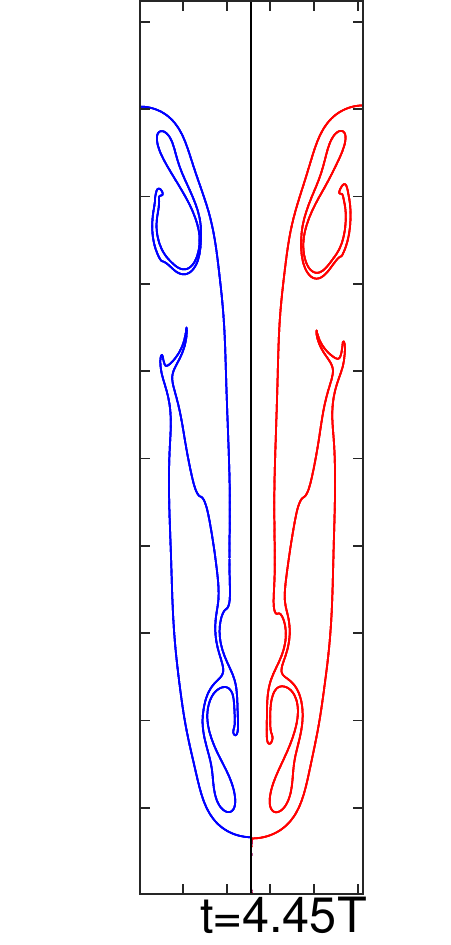}}\\
\subfloat[]{
    \includegraphics[width=0.2\textwidth,trim=145 20 145 23,clip]{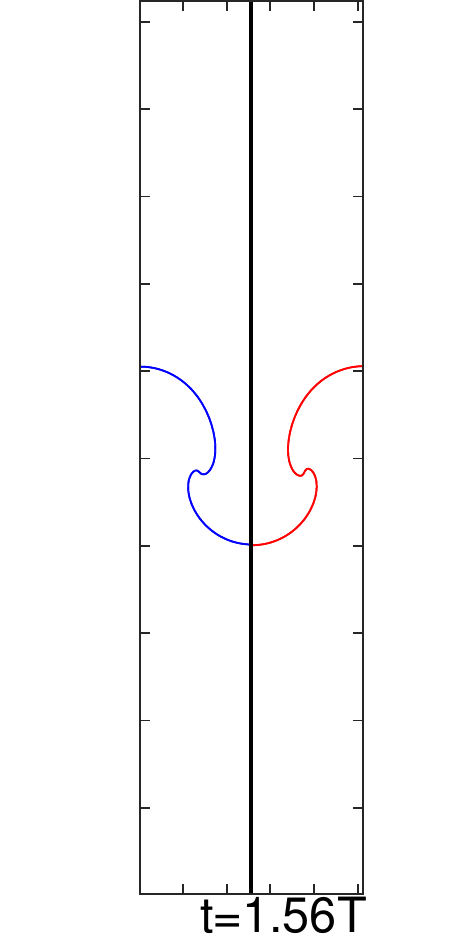}~
    \includegraphics[width=0.2\textwidth,trim=145 20 145 23,clip]{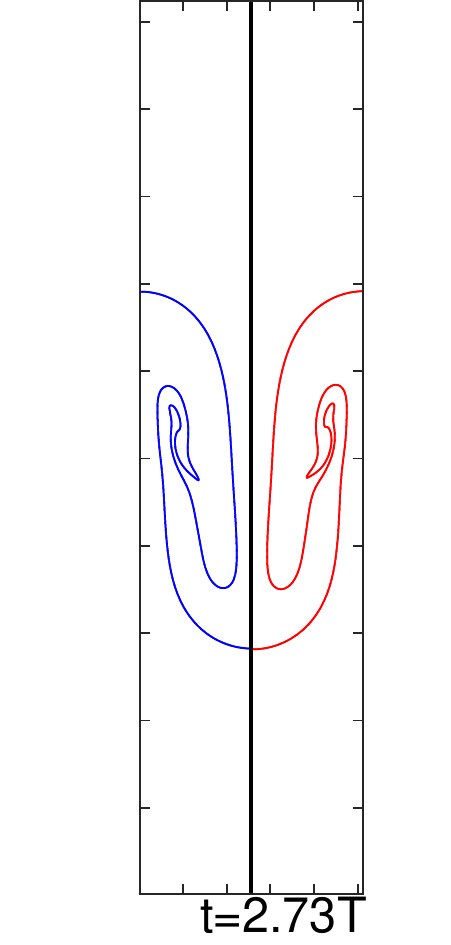}~
    \includegraphics[width=0.2\textwidth,trim=145 20 145 23,clip]{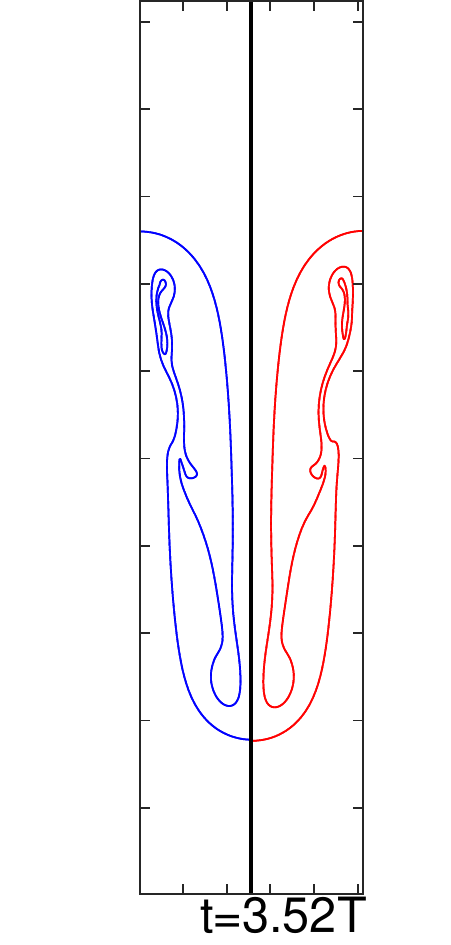}~
    \includegraphics[width=0.2\textwidth,trim=145 20 145 23,clip]{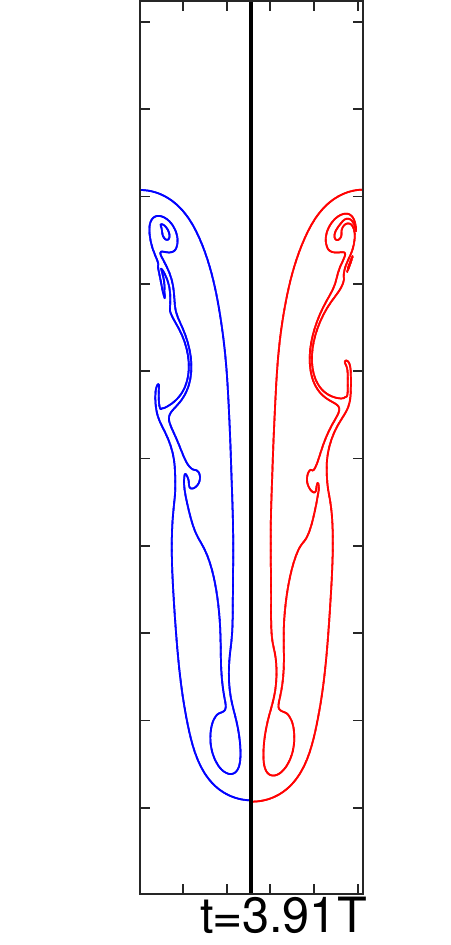}~
    \includegraphics[width=0.2\textwidth,trim=145 20 145 23,clip]{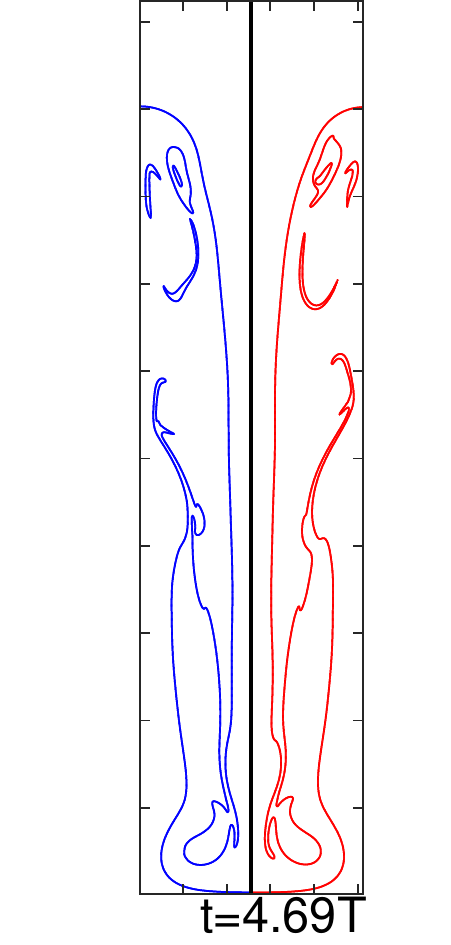}}
\caption{ Snapshots for a circular interface in the deformation field  at
(a) $A_t=0.1$ and (b) $A_t=0.25$. The time is normalized by the characteristic time $T=\sqrt{\lambda/Ag}$. The left column shows the results of the model in \cite{liang2014phase} and the right column shows the results of model II.}\label{case5PE1000Evo}
\end{figure}

All of the tests above are carried out by a given velocity field. In this case, flow instability of the Rayleigh-Taylor type is simulated in the rectangular domain $[0,d]\times[0,4d]$. Initially, a heavier fluid is placed on the top of a lighter one, and the interface between the two phase is perturbed by $\phi(x,y)=\tanh2(y-h)/D$, where $h=2d+0.05d\cos(2\pi x/\lambda)$ with $\lambda$ being the wavelength.
The periodic condition is applied on the side walls while the no-slip condition is imposed at the top and bottom walls.
 Due to the perturbation at the interface, the heavier fluid will  penetrate into the lighter fluid in a gravitational field. This problem is governed by two non-dimensional parameters: the Atwood number $A_t=(\rho_h-\rho_l)/(\rho_h+\rho_l)$ and
 the Reynolds number $Re=\lambda\sqrt{A_tg\lambda/(1+A_t)}/\nu$.
 In the simulations, the following parameters are used: $d=\lambda=256$,
  $\sqrt{g\lambda}=0.04$, $Re=400$, $A_t=0.1$, $\rho_h=1$, $D=4$, $Pe=1000$, $\tau_g=1$ and $\sigma=5.0\times 10^{-5}$. The viscosity ratio is 1.
Fig.~\ref{case5PE1000Evo} shows the interfacial evolution at $A_t=0.1$ and $A_t=0.25$. Since the interface shapes obtained by the model in \cite{liang2014phase} and model I are nearly the same, we only show the results of the model in \cite{liang2014phase}. It can be seen that the interface obtained by model II gets more stretched than those of model in \cite{liang2014phase} with the increase of the velocity.

\section{CONCLUSIONS}
In this paper, a high-order lattice Boltzmann model for the CHE is proposed.
First, through the Chapman-Enskog analysis, the source term on the convective time scale is defined as the relevant space derivative by coupling the NSEs.
The source term on the diffusive time scale is designed so that the CHE can be recovered  up to  the third order in terms of the expansion parameter $\epsilon$.
Then, we perform some tests to verify  the accuracy and stability of the present models. Numerical results show that
the source term expressed as the spatial derivatives may contribute to the stability
and accuracy of numerical calculation. For complex deformation fluid, the model II can capture the interface more accuracy than the model I because of considering the force term on the diffusive time scale. In addition, we have examined the effects of the Peclet number on the numerical results. It is found that both model I and model II have a good stability in a larger range of Peclet numbers.

\section*{ACKNOWLEDGEMENTS}
This study was supported by the National Key Research and Development Plan (Grant No. 2016YFB0600805).
\section*{References}
\bibliography{mybib}
\end{document}